\documentclass[twocolumn,preprintnumbers,pre]{revtex4-2}
\usepackage{graphicx, subfigure}
\usepackage{amssymb, amsmath,amssymb,amsfonts}
\usepackage{amsthm,mathrsfs,amsopn}
\usepackage{array}  
\usepackage{dcolumn}
\usepackage{bm}
\usepackage{enumerate}
\usepackage{color}
\usepackage[utf8]{inputenc}
\usepackage{mathtools}
\graphicspath{{./Figures/}}
\usepackage{hyperref}
\usepackage[normalem]{ulem}
\usepackage{tikz}
\usepackage{graphicx}
\usetikzlibrary{arrows.meta}
\usepackage{orcidlink}
\theoremstyle{plain} 

\begin{document}

\title{Collective decision-making with higher-order interactions on $d$-uniform hypergraphs}

\author{Thierry Njougouo$^{1,2}$\,\orcidlink{0000-0001-7706-7674}}
\email{thierry.njougouo@imtlucca.it}
\author{Timoteo Carletti$^{3}$\,\orcidlink{0000-0003-2596-4503}}
\author{Elio Tuci$^{4}$\,\orcidlink{0000-0001-7345-671X}}
\affiliation{$^1$IMT School for Advanced Studies\\ Piazza San Francesco 19, 55100 Lucca, Italy}
\affiliation{$^2$MoCLiS Research Group, Dschang, Cameroon.}
\affiliation{$^3$Department of Mathematics and Namur Institute for Complex Systems, naXys\\
Universit\'{e} de Namur, Rue Graf\'e 2, B5000 Namur, Belgium}
\affiliation{$^4$Faculty of Computer Science and Namur Institute for Complex Systems, naXys\\
Universit\'{e} de Namur, Rue Grandgagnage 21, B5000 Namur, Belgium}

\begin{abstract}
Understanding how group interactions influence opinion dynamics is fundamental to the study of collective behavior. In this work, we propose and study a model of opinion dynamics on $d$-uniform hypergraphs, where individuals interact through group-based (higher-order) structures rather than simple pairwise connections. Each one of the two opinions $A$ and $B$ is characterized by a quality, $Q_A$ and $Q_B$, and agents update their opinions according to a general mechanism that takes into account the weighted fraction of agents supporting either opinion and the pooling error, $\alpha$, a proxy for the information lost during the interaction. Through bifurcation analysis of the mean-field model, we identify two critical thresholds, $\alpha_{\text{crit}}^{(1)}$ and $\alpha_{\text{crit}}^{(2)}$, which delimit stability regimes for the consensus states. These analytical predictions are validated through extensive agent-based simulations on both random and scale-free hypergraphs. Moreover, the analytical framework demonstrates that the bifurcation structure and critical thresholds are independent of the underlying topology of the higher-order network, depending solely on the parameters $d$, i.e., the size of the interaction groups, and the quality ratio. Finally, we bring to the fore a nontrivial effect: the large sizes of the interaction groups, could drive the system toward the adoption of the worst option. 
\end{abstract}

\maketitle

\section{Introduction}
\label{sec:intro}
Understanding how interactions between individuals within a population lead to collective decision-making~\cite{bose2017collective} is one of the central challenges in complex systems. Opinion dynamics models provide a powerful conceptual framework for exploring these phenomena of collective decision-making, linking the microscopic mechanisms of social interaction to the macroscopic states of consensus, polarization, or coexistence of opinions~\cite{galam2002minority,galam2011sociophysics, lorenz2007continuous,ojer2025social,de2020emergence,castellano2009statistical}. These models have found a wide range of applications, from the diffusion of innovations~\cite{martins2009opinion,rogers2014diffusion}, rumors spreading~\cite{borge2012absence,castellano2009statistical}, epidemic spreading~\cite{pastor2015epidemic} and social behaviors~\cite{castellano2009statistical}, to the modeling of political opinion formation~\cite{friedkin2011social} and collective decision-making~\cite{hartnett2016heterogeneous,wen2024formulating,reina2024speed}. In all these application fields, the main question is how the structure of social interactions and the reliability of shared information jointly determine the accuracy and stability of collective decisions.

Most classical approaches describe social interactions as pairwise exchanges between connected individuals/agents, represented as links or edges in a network~\cite{sznajd2000opinion,dong2024evolution,hartnett2016heterogeneous,galam2011sociophysics,mengers2024leveraging}. It should be noted that, while such graph-based models have revealed essential mechanisms of consensus and polarization, they cannot fully capture the complexity of real-world scenarios where, for example, in many social and biological processes interactions occur in groups rather than dyads. More examples from daily life include, meetings, committee decisions, online discussions, and animal group choices, all usually involve groups interactions, where information is integrated and compared collectively and simultaneously among multiple participants.
Ignoring these group-level effects can lead to incomplete or even misleading predictions about emergent collective behavior. To address these limitations of the graph-based model, recent research has turned to richer representations, such as hypergraphs and simplicial complexes~\cite{zhang2023higher,battiston2020networks,petri2018simplicial}, which can model group exchanges as  higher-order interactions and more accurately reflect the group effects.

In the process of opinion exchange through social interactions, several mechanisms have been studied in the literature, among which the voter model~\cite{holley1975ergodic,sood2005voter}, where an agent adopts the opinion of a randomly chosen neighbor. In contrast, majority rule dynamics~\cite{galam2002minority,pereira2005majority,lambiotte2007majority} assume that individuals conform to the local majority within their neighbors, introducing nonlinear effects and allowing for abrupt transitions between consensus states~\cite{reina2024speed}. It remains important to note that, other extensions have included factors such as bounded confidence~\cite{hegselmann2015opinion}, or contrarian behavior~\cite{galam2004contrarian}, producing a wide variety of emergent outcomes.
Building upon the voter and majority models, recent studies have proposed a unifying framework that connects these two paradigms through the introduction of the called \textit{pooling error}~\cite{reina2024speed,njougouo2024impact}. The pooling error, regulates the accuracy with which agents perceive and aggregate the opinions of their neighbors.
Despite the success of these models, most of them are typically rooted on the assumption of pairwise interactions, leaving open the question of how higher-order, group-based interactions affect the dynamics of consensus formation and decision accuracy.
Some research works have tried to address this issue by extending opinion-based dynamics to higher-order interactions using hypergraphs or simplicial complexes. In these works, the opinion update process is driven by group-based rather than dyadic interactions. Examples include the work by Noonan and Lambiotte~\cite{noonan2021dynamics}, in which they propose a majority-rule dynamics on hypergraph and  show that even small interaction groups (e.g., triplets) can qualitatively modify the effective drift and the pathway to consensus compared with standard pairwise interactions. For voter-like dynamics, extension to adaptive hypergraph indicates that group influence can break magnetization conservation, amplifying small initial biases, and accelerating consensus~\cite{papanikolaou2022consensus}. In the same vein as voter-like dynamics, Golovin et al., in Ref.~\cite{golovin2024polyadic}, show that when the interaction structure coevolves over time, the coupling between group reconfiguration and opinion dynamics can lead to fragmentation or long-lived metastable states, depending on the group adoption rule on hypergraphs. Complementarily, polyadic voter variants such as the group-driven voter model developed in Ref.~\cite{kim2025competition} highlight how the interplay between group-size constraints and nonlinear update rules can lead to optimal consensus times, thereby providing a concrete mechanism through which higher-order interactions qualitatively affect classical voter phenomenology. These findings suggest that the \textit{pooling error}~\cite{reina2024speed,njougouo2024impact} mechanism---which controls how accurately agents aggregate neighbors' opinions---may interact nontrivially with group-based information pooling, strengthening the motivation for the higher-order perspective adopted here. Finally, this point of view also aligns with the literature on \emph{competing spreading processes} on higher-order structures, developed in epidemic settings~\cite{li2022competing,nie2022markovian} and, more recently, in collective decision-making models with multiple mutually exclusive options, where higher-order social influence can break symmetry and resolve decision deadlocks~\cite{march2026symmetry}.

In this work we focus on group interactions in the context of the Best-of-$n$ problem~\cite{valentini2017best}, which formalizes how a population  of agents, faced with several alternatives of different qualities, can (or cannot) converge to the best one through decentralized interactions and information exchange~\cite{bose2017collective,mann2018collective,pratt2002quorum,seeley2006group,hartnett2016heterogeneous}. The Best-of-$n$ problem underlies many natural and artificial systems, from nest-site selection in honeybees and ants~\cite{seeley2006group,pratt2002quorum} to human decision-making under uncertainty~\cite{marshall2009optimal}. Recent studies have shown that the structure of interactions — network topology, connectivity, or group size — can strongly influence the ability of a collective to identify the optimal option~\cite{reina2024speed,njougouo2024impact,moeinifar2021zealots}. 
Building on this line of research, the main contribution of this work is to extend the Best-of-$2$ problem, traditionally studied under the assumption of purely pairwise interactions, to the case of higher-order interactions on $d$-uniform hypergraphs of random and scale-free types. This investigation highlights how higher-order interactions shape opinion dynamics on uniform hypergraphs by showing that the size of interaction groups and the quality ratio of competing options determine the regimes of consensus and opinion stability, independently of the underlying network structure. Moreover our analysis allows to emphasize the impact of large size groups, that can drive the system toward the adoption of the worst option. 

Our results are based on the introduction and use of the Heterogeneous Mean Field adapted to the higher-order networks setting, allowing us to derive an Ordinary Differential Equation governing the evolution of the (weighted) average fraction of agents supporting options $A$ and $B$. Having set this framework, we can analytically study the role of the involved model parameters and higher-order network structure, and obtain bifurcation diagrams to explain the system behavior. Eventually, we complemented the results obtained from the analytical work, by developing an Agent Based Model and by numerically integrating its solutions.

The remainder of this paper is organized as follows. 
Sec.~II introduces the model and the opinion exchange mechanism. 
Sec.~III presents the Heterogeneous Mean-Field approximation used to construct the analytical framework and displays the results of a bifurcation analysis, comparing theoretical predictions with Agent Based simulation results. 
Finally, Sec.~IV presents the conclusions, discusses the implications of our findings, and outlines directions for future research.

\section{The model}
\label{sec:themodel}
Let us consider a population composed of $N$ agents aimed at performing a choice among two options, $A$ and $B$, each endowed with a given quality, $Q_A>0$ and $Q_B>0$; without loss of generality, we will assume $Q_B\leq Q_A$, hence the quality ratio $Q=Q_B/Q_A$ is bounded to $[0,1]$. This quality ratio quantifies the difficulty of the decision problem. Therefore, low $Q$ values indicate an easy discrimination between options, while a $Q \to 1$ corresponds to a hard decision task in which both options have nearly equal quality.
Agents experience group interaction and change their opinions depending on the votes expressed by their peers. We use hypergraph to encode group interactions: each agent is represented by a node, and each hyperedge defines a group, i.e., the set of agents jointly participating in the voting process (either changing or maintaining their opinion). Although agents interact in groups, each agent contribution to the group is linearly proportional to the quality of her opinion. More precisely, we consider a hypergraph $H=(V,E)$ with $N=|V|$ nodes 
and $L=|E|$ hyperedges, where each hyperedge $e_j \subseteq V$ has size 
$d_j = |e_j|$, for $j=1,\dots,L$. In general hyperedge sizes follow a distribution 
$s:\mathbb{N}\rightarrow \mathbb{R}_+$, where $s_d$ is the probability that 
a randomly selected hyperedge has size $d$, with $\sum_{d\geq1}s_d=1$. In the $d$-uniform case, which is the main setting considered here, all 
hyperedges have size $d$, so that $s_d=1$ and $s_{d'}=0$ for $d'\neq d$.
Let us observe that we hereby assume that an hyperedge of size $d$ contains $(d+1)$ nodes~\cite{muolo2024phase}, this is the usual definition in the case of simplicial complexes but it is often used also in the case of hypergraph, moreover it will allow to slightly simplify our notations. Based on this definition, we are thus excluding hyperedges containing a single node, i.e., $s_0=0$. Given an agent, say $i$, we will define her hyperdegree, $h_i$, to be the number of hyperedges incident to $i$, namely the number of hyperedges to which $i$ belongs. We assume the hyperdegrees to be distributed according to a function 
$p:\mathbb{N}\rightarrow \mathbb{R}_+$, where $p_k$ denotes the probability that a randomly selected agent has hyperdegree $h_i=k$, with $\sum_{k\geq 1} p_k = 1$. By analogy with the case of graphs, the choice of the distribution $p_k$ determines the structural class of the resulting hypergraph. For instance, 
if $p_k$ follows a Poisson distribution, the corresponding structure can be interpreted as a random hypergraph, in analogy with Erd\H{o}s--R\'enyi random graphs~\cite{erdds1959random}. 
On the other hand, if $p_k$ follows a power-law distribution 
$p_k \sim k^{-\gamma}$ with exponent $2 < \gamma < 3$, the resulting hypergraph displays scale-free properties, similarly to scale-free networks observed in many real systems~\cite{barabasi1999emergence,jhun2019simplicial}. Such heterogeneous higher-order structures are commonly studied within the framework of configuration models for hypergraphs and higher-order networks~\cite{courtney2016generalized}. Throughout this work, scale-free hypergraphs are generated with a power-law exponent fixed to $\gamma = 2.5$, similarly to the framework considered in~\cite{courtney2016generalized}, and as detailed in Appendix.~\ref{sec: appps1}.

After constructing the network and specifying the initial conditions (randomly assigned with equal probability to the two options in this study), the system evolves according to the following rules.\\
i) Time evolves by discrete steps and each time step an agent is randomly selected with uniform probability;\\
ii) Let the selected agent to be the $i$-th one, assume her hyperdegree to be $k$, i.e., $h_i=k$, and for a sake of definitiveness assume she supports option $A$. Then, she selects one among the incident hyperedges according to some function of the hyperedge size, $g_k(d)$, with $\sum_{j=1}^k g_k(d_j)=1$. Let us observe that in this way the selection depends on the size of the hyperedges and thus we can introduce a bias with respect to the latter; namely one can favor hyperedges with larger size or do the contrary. We could also consider more tricky behavior, where the preferred size depends on the hyperdegree. On the other hand if $g_k(d_j)=1/k$, then the selection process is uniformly at random.
Unless stated otherwise, throughout this paper we assume that the incident hyperedge is selected uniformly at random, independently of its size.\\
iii) Let the selected hyperedge to be $e_r$ with size $d_r$, i.e., there are $(d_r+1)$ agents, one of which is the focal agent $i$. The agents in the group produce two signals
\begin{equation}
\label{eq:defnBdies}
n_{A,r}^\# = \frac{n_{A,r}}{n_{A,r}+Q n_{B,r}} \text{ and } 
n_{B,r}^\# = \frac{Q n_{B,r}}{n_{A,r}+Q n_{B,r}}\, ,
\end{equation}
namely the weighted, with respect to quality ratio $Q$, number of agents supporting option $A$, $n_{A,r}$, and $B$, $n_{B,r}$, in the group, excluding the focal agent. We thus have $n_{A,r}^\#+n_{B,r}^\#=1$. Then with probability $P_\alpha(n_{B,r}^\#)$ the focal agent changes her option and passes to $B$. Here the function $P_\alpha(x)$ is some nonlinear function encoding for the microscopic rule determining the agents behavior, captured by the parameter $\alpha$, but it can also depend on the hyperedge size, $d_r$. One possible choice could be \cite{reina2024speed}:
\begin{equation}
\label{eq:Px}
 P_\alpha(x) = 
            \begin{dcases}
             \frac{1}{2}-\frac{1}{2}\left(1-2x\right)^{\alpha} & \text{if $ 0\leq x \leq \frac{1}{2}$}\\
              \frac{1}{2}+\frac{1}{2}\left(2x-1\right)^{\alpha}  & \text{if $ \frac{1}{2} < x \leq 1$}\,,
            \end{dcases}\, 
\end{equation}
where $x \in [0,1]$ represents the weighted proportion of agents holding a different opinion, and the parameter $\alpha \geq 0$ quantifies the pooling error~\cite{reina2024speed,njougouo2024impact}, i.e., the inaccuracy with which an agent perceives the  opinions of her neighbors. The value of $\alpha = 0$ corresponds to a decision  taken according to the majority model, indeed $P_{0}(1)=1$ if $x>1/2$ and $0$ otherwise, the value of $\alpha = 0$ is thus associated with no pooling error: the focal agent is capable to correctly gather the whole available information. By increasing $\alpha$, the pooling error also increases, indeed for $\alpha = 1$, the function $P_1(x)$ is linear and the focal agent selects according to the voter model strategy, i.e., she picks one neighbor opinion at random. Very large $\alpha$ are associated to random choices because $P_\alpha(x)$ tends to the constant value $1/2$.\\
iv) The focal agent $i$ is inserted back into the system with a possible new opinion, time advances by one step and the process goes back to i). Let us conclude to observe that if the focal agent supported option $B$, then she would have changed her mind with probability $P_\alpha(n_{A,r}^\#)=1-P_\alpha(n_{B,r}^\#)$.

Let us conclude this section with the following remark. In Ref.~\cite{reina2024speed,njougouo2024impact} a similar model to study the best-of-$2$ problem has been proposed in the framework of pairwise interactions, namely agents are assumed to be nodes of a network and links describe  social tights. Despite the similarity in the process, the hypergraph framework changes the effective amount of social information used at each update; while pairwise updates involve all the neighbors of a given focal agent, here the latter updates according to one selected incident hyperedge. Thus, the number of peers involved in a single update is controlled by the size of the selected hyperedge, rather than by the total number of neighbors across all groups. In this way we can encode the fact that individuals are part of different social groups, hyperedges, and information flows inside each group.

\section{The hypergraph Heterogeneous Mean Field}
\label{sec:HHMF}
To make a step further in the understanding of the system, we assume that all nodes with the same hyperdegree exhibit the same behavior, this a sort of hypergraph version of the Heterogeneous Mean Field (HMF) hypothesis \cite{colizza2007reaction, pastor2015epidemic, costa2022heterogeneous}. Therefore, we define $A_k$ (resp. $B_k$) as the number of agents with hyperdegree $k$ and opinion $A$ (resp. opinion $B$), and $N_k$ as the total number of agents with hyperdegree $k$. Hence, $A_k+B_k=N_k$ for all $k$. For the sake of normalization and simplification, we also define $a_k=A_k/N_k$ and $b_k=B_k/N_k$ as the fractions of agents with opinion $A$ and $B$, respectively,
with hyperdegree $k$.

To describe the time evolution of $a_k$ for a generic $k$ we can observe that it increases when an agent with hyperdegree $k$ and opinion $B$ changes her opinion to $A$, or it decreases when an agent with hyperdegree $k$ and opinion $A$ changes her opinion to $B$. To compute the frequency of these events, we start by computing $n_{A,r}^\#$ and $n_{B,r}^\#$ under the HMF assumption. Let us consider again the focal agent $i$ to have hyperdegree $k$, let moreover assume $e_r$ to be the selected incident hyperedge with size $d_r$. In this group there could be a number $\omega$, with $\omega=0,\dots, d_r$, of agents supporting option $B$, and thus $(d_r-\omega)$ in favor of option $A$.

\begin{figure}[htp!]
    \centering
        \includegraphics[width=0.5\textwidth]{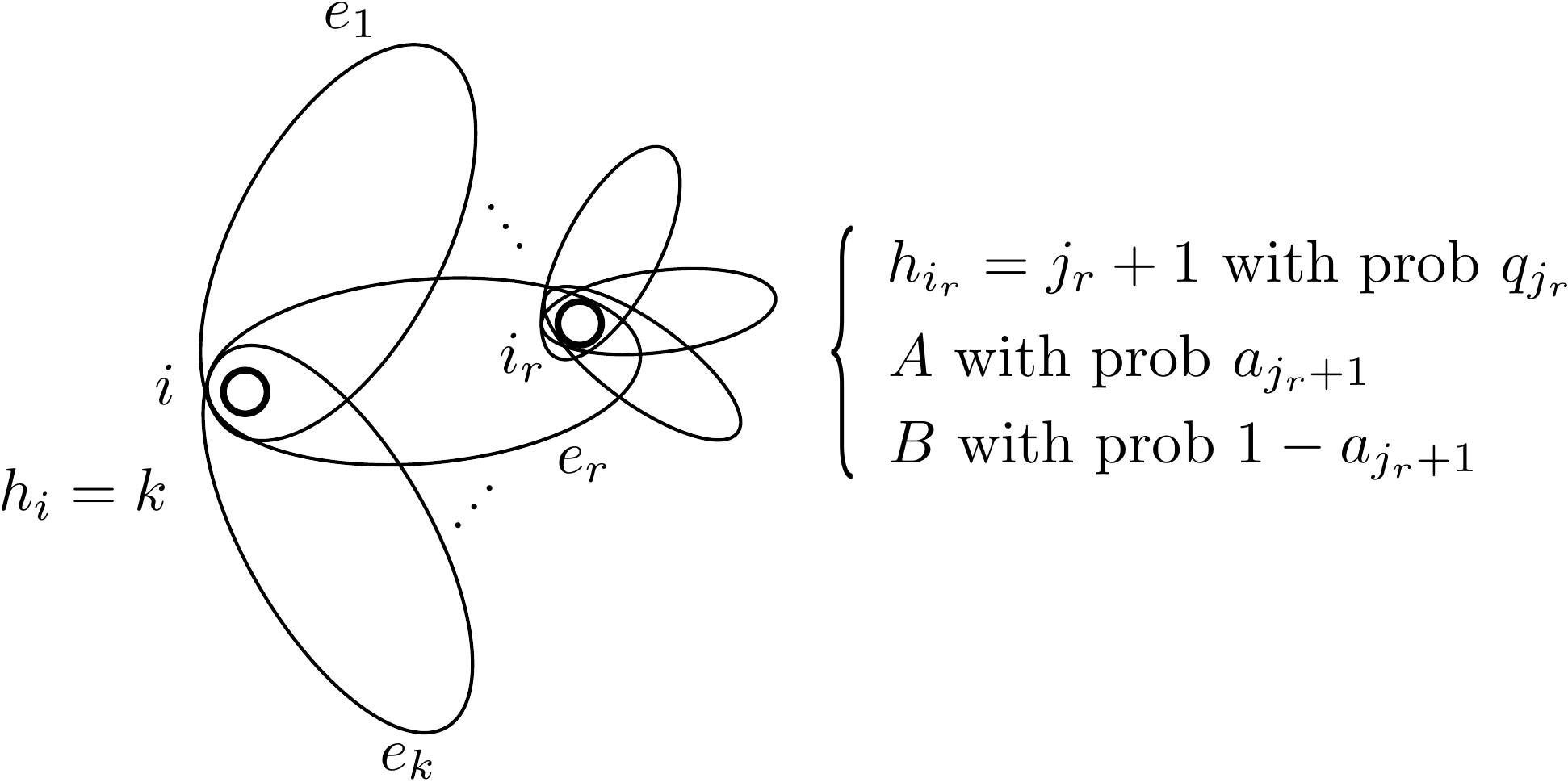} 
        \caption{Schematic representation of the probabilities involved in the hypergraph heterogeneous mean-field computations. The focal node $i$ has $k$ incident hyperedges (hyperdegree $k$), denoted by $e_1,\dots,e_r,\dots,e_k$. A generic node $i_r$ in the hyperedge $e_r$ has an excess hyperdegree $j_r$ with probability $q_{j_r}$ and therefore hyperdegree $j_r+1$. With probability $a_{j_r+1}$, she has opinion $A$, and therefore with probability $1-a_{j_r+1}$, she has opinion $B$.}
        \label{fig:HHMFscheme}
\end{figure}

Because we are interested into the classes, $a_{k'}$ and $b_{k'}$, we have to compute the probability that the above nodes with opinion $A$ and $B$ also have hyperdegree $k'$, for some $k'\geq 1$. To do this we recur to the concept of {\em excess degree}~\cite{newman2018networks} but in the hypergraph setting; more precisely we define the {\em excess hyperdegree} to be the number of incident hyperedges that a neighbor of the focal agent has, without considering the selected hyperedge $e_r$ containing also the focal agent. In other words, the excess hyperdegree of an agent is her hyperdegree minus the selected hyperedge (i.e., minus one) as illustrated in  Fig.~\ref{fig:HHMFscheme}, stated differently, the excess hyperdegree counts the number of different hyperedges to which an agent can communicate her opinion, without taking into account the hyperedge from where she received the information.\\
Letting $p_k$ to be the probability that a generic node has hyperdegree $k$, and $q_k$ the probability for a node to have excess hyperdegree equal to $k$, then
\begin{equation}
\label{eq:excess}
q_k = \frac{(k+1)p_{k+1}}{\langle k \rangle} \quad \forall k\geq 0\, ,
\end{equation}
where $\langle k \rangle=\sum_k kp_k$ is the average node hyperdegree, and trivially $\sum_k q_k=1$. 

Let us consider a generic focal agent $i$ with hyperdegree $k$ and opinion $B$ and assume she selects the hyperedge $e_r$ (see Fig.~\ref{fig:HHMFscheme}). Let $q_{j_r}$ be the probability that a generic agent $i_r$ in the hyperedge $e_r$ has excess hyperdegree $j_r\geq 0$. Let $a_{j_r+1}$ be the probability that agent $i_r$ has opinion $A$, and $b_{j_r+1}=1-a_{j_r+1}$ the probability she has opinion $B$. By considering all the $d_r$ agents belonging to the hyperedge $e_r$ we can conclude that $q_{j_1}\dots q_{j_{d_r}}$ determine the joint probability that each agent has excess hyperdegree $j_1,\dots,j_{d_r}$. We can then define $\pi_{d_r,\omega}$ as the probability that $\omega$ agents among the $d_r$ in the hyperedge $e_r$ have opinion $B$ and thus $d_r-\omega$ agents have opinion $A$. Therefore, the term $\pi_{d_r,\omega}$ is a linear combination of the products of $a_{j_m+1}$ and $(1-a_{j_m+1})$, with $m=1,\dots,d_r$. Hence, we can compute the weighted proportion of agents with opinion $A$ as $n_{A,r}^\#= (d_r-\omega)/[d_r-\omega+Q\omega]$, and this event occurs with probability $q_{j_1}\dots q_{j_{d_r}}\pi_{d_r,\omega}$. In conclusion, agent $i$ with opinion $B$ can change opinion with probability 
\begin{equation*}
    P_{B\rightarrow A} \propto q_{j_1}\dots q_{j_{d_r}}\pi_{d_r,\omega}P_\alpha\left(\frac{d_r-\omega}{d_r-\omega+Q\omega}\right)\, ,
\end{equation*}
We refer the readers to Ref.~\cite{reina2024speed}, where a similar and 
more detailed study was carried out for graphs. The same methods and 
techniques are used here to obtain the analytical model. To avoid repetition, 
we will omit some of the intermediate steps, but they can be consulted in 
the aforementioned reference.

In a similar way, we can compute the decrease rate of agents with hyperdegree $k$ and opinion $A$. In this case the probability $P_{A\rightarrow B}$ will depend on $P_\alpha\left(\frac{Q\omega}{d_r-\omega+Q \omega}\right)$, that is, the weighted proportion of agents with opinion $B$ assuming that $\omega$ agents among them have opinion $B$. Note that, because $P_\alpha(n_{A,r}^\#)+P_\alpha(n_{B,r}^\#)=1$, we have that 
\begin{eqnarray*}
P_\alpha\left(\frac{Q\omega}{d_r-\omega+Q \omega}\right) &=&1-P_\alpha\left(1-\frac{Q\omega}{d_r-\omega+Q \omega}\right) \\
&=& 1- P_\alpha\left(\frac{d_r-\omega}{d_r-\omega+Q \omega}\right)\, .
\end{eqnarray*}
Thus, within the HMF  approximation introduced above, 
for each $k$ the evolution of the fraction of agents holding opinion $A$, denoted 
by $a_k$, can be similarly adapted for opinion $B$ using the relation $a_k + b_k = 1$, 
i.e., $b_k = 1 - a_k$. 
To determine the evolution of $a_k$ we have to take into account the probability of selecting the hyperedge $e_r$ with size $d_r$. Thus, we can wrap together the above expressions and obtain the time evolution of $a_k$ for a generic degree $k$:
\begin{widetext} 
\begin{eqnarray}
\label{eq:dakdt}
\frac{da_k}{dt} &=& (1-a_k)\sum _{r=1}^kg_k(d_r)\sum_{j_1,\dots,j_{d_r}}q_{j_1}\dots q_{j_{d_r}} \sum_{\omega=0}^{d_r} \pi_{d_r,\omega}P_\alpha\left( \frac{d_r-\omega}{d_r-\omega +\omega Q}\right) \notag\\
&-&a_k\sum _{r=1}^kg_k(d_r)\sum_{j_1,\dots,j_{d_r}}q_{j_1}\dots q_{j_{d_r}} \sum_{\omega=0}^{d_r} \pi_{d_r,\omega}\left[1- P_\alpha\left( \frac{d_r-\omega}{d_r-\omega +\omega Q}\right)\right]\,.
\end{eqnarray}
\end{widetext} 

For the sake of pedagogy, let us explain each term on the right-hand side. The leftmost term in the first line, $(1-a_k)$, is the probability that the focal agent has hyperdegree $k$ and does not have opinion $A$, she hence has opinion $B$. The term $g_k(d_r)$ is the probability to select the hyperedge $e_r$ with size $d_r$ among the $k$ possible ones; then the product $q_{j_1}\dots q_{j_{d_r}}$ evaluates the joint probability that each node belonging to the hyperedge $e_r$ has excess degree $j_1,\dots,j_{d_r}$. The sum $\sum_{j_1,\dots,j_{d_r}}$ allows to consider all the possibilities. For a given choice of $j_1,\dots,j_{d_r}$, the next term, $\pi_{d_r,\omega}$, determines the probability that $\omega$ nodes among the $d_r$ ones have opinion $B$ and thus $d_r-\omega$ have opinion $A$. The sum $\sum_{\omega=0}^{d_r}$ allows to consider all the possibilities from $\omega=0$, all agents have opinion $A$, i.e., to $\omega=d_r$, i.e., all agents have opinion $B$. Finally, the term $P_\alpha\left( \frac{d_r-\omega}{d_r-\omega +\omega Q}\right)$ is the probability the focal agent with opinion $B$ changes her mind because there are $(d_r-\omega)$ agents with opinion $A$ and $\omega$ agents with opinion $B$. The remaining terms denote the opposite process where the selected agent has opinion $A$ and hyperdegree $k$, with probability $a_k$, and she changes opinion after an interaction with her neighbors with opinion $B$.

Although Eq.~\ref{eq:dakdt} describes the evolution of the fraction of agents holding opinion $A$ for each hyperdegree $k$, analyzing the collective dynamics by computing the equilibria directly from this equation remains challenging. To bridge this gap, we consider the overall network behavior through the mean fraction of agents with opinion $A$ and excess hyperdegree $q_j$, which allows us to connect individual hyperdegree dynamics to the global network behavior:
\begin{equation}
\langle a \rangle := \sum_{j \geq 0} q_j a_{j+1}.
\label{eq:am}
\end{equation}
Let us use Eq.~\ref{eq:dakdt} to determine the time evolution of $\langle a \rangle$; some straightforward computations (reported in Appendix~\ref{sec:app1}) allow to get Eq.~\ref{eq:edomain}
\begin{widetext} 
\begin{eqnarray}
\label{eq:edomain}
\frac{d\langle a \rangle}{dt} = -\langle a \rangle+\sum_k q_k \sum _{r=1}^{k+1}g_{k+1}(d_r) \sum_{\omega=0}^{d_r} \binom{d_r}{\omega} \langle a\rangle^{d_r-\omega}\left(1-\langle a\rangle\right)^\omega P_\alpha\left( \frac{d_r-\omega}{d_r-\omega +\omega Q}\right)\, .
\end{eqnarray}
\end{widetext} 
This model describes the evolution of the weighted mean fraction of agents with opinion $A$ on a given hypergraph. It is also applicable to heterogeneous hypergraphs with variable hyperedge sizes, but the resulting Eq.~\ref{eq:edomain} is analytically cumbersome. A simpler, analytically tractable case is that of $d$-uniform hypergraphs, where all hyperedges have the same size.

\subsection{Case of a $d$-uniform hypergraph model}
\label{sec:dhyper}

For the sake of pedagogy and to simplify the analysis, let us consider the case of the $d$-regular hypergraph, i.e., when all the hyperedges have the same size, $d$, and thus each one contains $(d+1)$ nodes. Because there are not differences among hyperedges we also assume the selection of the hyperedge to be uniformly random, namely we set $g_k(d)=1/k$. Then Eq.~\ref{eq:edomain} rewrites
\begin{eqnarray*}
\frac{d\langle a \rangle}{dt} &=& -\langle a \rangle+\sum_k q_k \sum _{r=1}^{k+1}\frac{1}{k+1} \sum_{\omega=0}^{d} \binom{d}{\omega} \langle a\rangle^{d-\omega}\left(1-\langle a\rangle\right)^\omega \times \\ &\times& P_\alpha\left( \frac{d-\omega}{d-\omega +\omega Q}\right)\, .
\end{eqnarray*}
Let us observe that the sum over $\omega$ is now independent from $r$ and can thus be factorized, moreover $\sum _{r=1}^{k+1}\frac{1}{k+1}=1$ and $\sum_k q_k=1$, and we eventually get
\begin{align}
\label{eq:mainedod}
\frac{d\langle a \rangle}{dt} &= -\langle a \rangle 
+ \sum_{\omega=0}^{d} \binom{d}{\omega} \langle a\rangle^{d-\omega} \left(1-\langle a\rangle\right)^\omega \notag \\
&\quad \times P_\alpha\left( \frac{d-\omega}{d - \omega + \omega Q} \right) 
\equiv f_\alpha(\langle a\rangle)\, .
\end{align}

To characterize the long-term behavior of the system, we first determine the equilibria of Eq.~\ref{eq:mainedod} and then, examine the stability properties of these equilibria by analyzing how they depend on the model parameters. The equilibrium points of the system correspond to the stationary states, which are obtained by solving $f_\alpha(\langle a^* \rangle)=0$, i.e., by requiring the time derivatives to vanish. A straightforward analysis shows that $f_\alpha(0)=f_\alpha(1)=0$, implying that $\langle a^* \rangle = 0$ and $\langle a^* \rangle = 1$ are solutions of the equation $f_\alpha(\langle a^* \rangle)=0$. These solutions therefore correspond to equilibrium points of the system defined by Eq.~\ref{eq:mainedod}, whose stability depends on the model parameters. When the equilibrium $\langle a^* \rangle = 1$ is stable, where $a_k=1$ for all $k$, the system reaches a consensus decision in favor of option $A$, meaning that all agents hold opinion $A$ and the population correctly selects the best option, since $Q_A > Q_B$. Conversely, when the equilibrium $\langle a^* \rangle = 0$ is stable,  where $a_k=0$ for all $k$, the system converges to a consensus decision in favor of option $B$, with all agents eventually adopting opinion $B$. In this case, the population makes a collective error by selecting the option with inferior quality ($Q_B < Q_A$). For certain ranges of the parameters $\alpha$, $Q$, and $d$, a third nontrivial equilibrium $0 < \langle \tilde{a}^* \rangle < 1$ may exist. The analytical expression of this equilibrium is difficult to obtain due to the complexity of the underlying equation, and it is therefore evaluated numerically. This equilibrium corresponds to a polarized state in which agents holding opinions $A$ and $B$ coexist without reaching consensus, leading to a collective decision deadlock.
 
To study the system fate we then analyze the stability of those equilibria.

\subsubsection{ Stability analysis}
\label{sec:dhyper}
The stability of the three equilibria described above can be assessed by analyzing the sign of the derivative of $f_\alpha(\langle a \rangle)$ evaluated at the equilibrium points. Using the properties $P_\alpha(0)=0$ and $P_\alpha(1)=1$, which hold for all values of $\alpha$, the expression of $f_\alpha(\langle a \rangle)$ can be simplified. In particular, for the first term of the sum, corresponding to $\omega = 0$, one has
\begin{equation*}
    P_\alpha\left(\frac{d-\omega}{d-\omega+\omega Q}\right)
    = P_\alpha\left(\frac{d}{d}\right)
    = P_\alpha(1),
\end{equation*}
so that the corresponding contribution reduces to $\langle a^* \rangle^{d}$. Similarly, for the last term of the sum, corresponding to $\omega = d$, one finds

\begin{equation*}
    P_\alpha\!\left(\frac{d-\omega}{d-\omega+\omega Q}\right)
    = P_\alpha\left(\frac{0}{dQ}\right)
    = P_\alpha(0),
\end{equation*}
and the associated contribution vanishes. Taking these two contributions into account, the expression of $f_\alpha(\langle a \rangle)$ can be rewritten by restricting the summation to $1 \leq \omega \leq d-1$ as:
\begin{align}
\label{eq:f1}
f_\alpha(\langle a\rangle)
&= -\langle a\rangle
+ \langle a\rangle^{d}
+ \sum_{\omega=1}^{d-1}
\binom{d}{\omega}
\langle a\rangle^{d-\omega}
\left(1-\langle a\rangle\right)^{\omega}\notag \\
&\quad \times P_\alpha\!\left(\frac{d-\omega}{d-\omega+\omega Q}\right).
\end{align}

From Eq.~\ref{eq:f1}, it follows directly that $f_\alpha(0)=f_\alpha(1)=0$, independently of the values of the model parameters $\alpha$, $Q$ and $d$. As a consequence, the states $\langle a^* \rangle = 0$ and $\langle a^* \rangle = 1$ are always equilibrium points of the system. Depending on the values of the model parameters and on the stability of these equilibria, they may correspond to full consensus on options $B$ and $A$, respectively. To determine which of these outcomes is realized, we now proceed to analyze the stability of these trivial equilibrium points.

Let us observe that each $d$-hyperedge is composed by $(d+1)$ nodes, thus $d=1$ returns the network case. In the following we will hence assume $d\geq 2$, to deal with hypergraphs. Let us consider separately the case $d=2$ and $d\geq 3$. In the first case we have
\begin{equation*}
f_\alpha(\langle a\rangle) = -\langle a\rangle+ \langle a\rangle^2 + 2 \langle a\rangle\left(1-\langle a\rangle\right) P_\alpha\left( \frac{1}{1+ Q}\right)\, ,
\end{equation*}
and thus, the derivative of $f_\alpha(\langle a \rangle)$ with respect to $\langle a \rangle$ can therefore be expressed as
\begin{equation*}
f_\alpha^\prime(\langle a\rangle)=-1+ 2\langle a\rangle+2 \left(1-2\langle a\rangle\right) P_\alpha\left( \frac{1}{1+ Q}\right)\, .
\end{equation*}
\\
Let us observe that $Q<1$ and thus ${1}/{(1+ Q)}<1/2$, that implies $P_\alpha\left( \frac{1}{1+ Q}\right)>1/2$ for all $\alpha$. We can thus conclude that for all $\alpha\geq 0$
\begin{eqnarray*}
f_\alpha^\prime(0)=-1+2 P_\alpha\left( \frac{1}{1+ Q}\right) >0 \\ 
\text{and} \quad 
f_\alpha^\prime(1)= 1-2P_\alpha\left( \frac{1}{1+ Q}\right)<0\, .
\end{eqnarray*}
namely $\langle a^*\rangle=0$ is always unstable while $\langle a^*\rangle=1$ is always stable: the population should always converge to all-$A$, if the initial population contains at least one agent supporting option $A$. No other equilibria are allowed.

Let us now consider the case $d\geq 3$. From Eq.~\eqref{eq:f1} we get the general form of the first derivative,
\begin{widetext} 
\begin{equation}
\label{eq:fprime1}
f_\alpha^\prime(\langle a\rangle)=-1+ d\langle a\rangle^{d-1}+\sum_{\omega=1}^{d-1} \binom{d}{\omega} \langle a\rangle^{d-\omega-1}\left(1-\langle a\rangle\right)^{\omega -1}\left[(d-\omega)-d\langle a\rangle\right]P_\alpha\left( \frac{d-\omega}{d-\omega +\omega Q}\right)\, ,
\end{equation}
\end{widetext} 
Let us now investigate the sign of the first derivative $f_\alpha^\prime(\langle a \rangle)$ evaluated at the equilibrium points $\langle a^* \rangle = 0$ and $\langle a^* \rangle = 1$, as follows:

\begin{eqnarray}
\label{eq:fprime10}
f_\alpha^\prime(0)&=&-1+ d P_\alpha\left( \frac{1}{1+(d-1) Q}\right)\\
\label{eq:fprime11}
f_\alpha^\prime(1)&=&
-1+dP_\alpha\left( \frac{Q}{d-1 + Q}\right)\, ,
\end{eqnarray}
where we used the property $1-P_\alpha(\langle a \rangle)=P_{\alpha}(1-\langle a \rangle)$ to rewrite the last term.

Given the importance of the sign of the derivative $f_\alpha'(\langle a \rangle)$ at $\langle a \rangle=0$ and $\langle a \rangle=1$ for the stability of the equilibrium points $0$ and $1$, respectively, we analyze $f_\alpha'(0)$ by considering two cases: $(d-1)Q < 1$ and $(d-1)Q \geq 1$.
In the first case we have $\frac{1}{1+(d-1) Q} > \frac{1}{2}$ and thus $P_\alpha\left( \frac{1}{1+(d-1) Q}\right)>1/2>1/d$, where the last inequality follows from $d\geq 3$. We can thus conclude that $\langle a^* \rangle = 0$ is unstable for all $\alpha\geq 0$ if $(d-1) Q< 1$. Let us now consider $(d-1) Q\geq 1$, hence $\frac{1}{1+(d-1) Q} \leq \frac{1}{2}$ and thus Eq.~\ref{eq:fprime10} rewrites (see Eq.~\ref{eq:Px})
\begin{equation}
\label{eq:eqfpo}
 f_\alpha^\prime(0) = 
 -1+\frac{d}{2}\left[1-\left(\frac{(d-1)Q-1}{(d-1) Q+1}\right)^\alpha\right]\, ,
\end{equation}
there is thus a critical value of $\alpha$ for which $f^\prime(0)=0$.  This critical value is obtained by solving the condition $f_\alpha'(0)=0$, where the expression of $f_\alpha'(0)$ is given by Eq.~\ref{eq:eqfpo}.
\begin{equation}
\alpha_{crit}^{(1)}=\frac{\log\left(\frac{d-2}{d}\right)}{\log\left(\frac{(d-1)Q-1}{(d-1) Q+1}\right)}\, .
 \label{eq:eqalc1}
\end{equation}
It follows that $f_\alpha'(0) < 0$ $\forall \alpha < \alpha_{\rm crit}^{(1)}$. In other words, if $(d-1)Q \geq 1$, the all-$B$ solution (i.e., $\langle a^* \rangle = 0$ and then, no agent supporting $A$), in which the population collectively selects the inferior option, is stable provided that $\alpha < \alpha_{\rm crit}^{(1)}$.

Let us now consider the stability of the equilibrium all-$A$, i.e., $\langle a^* \rangle = 1$. Because $d-1+Q>2$ we get $Q/[d-1+Q]<1/2$ and thus from Eq.~\eqref{eq:fprime11} we get
\begin{equation*}
f_\alpha^\prime(1)=
 -1+\frac{d}{2}\left[1-\left(\frac{d-1-Q}{d-1 + Q}\right)^\alpha\right]\, .
\end{equation*}
There is thus a second critical value of $\alpha$ for which $f^\prime(1)=0$, whose explicit value is 
\begin{equation}
\alpha_{crit}^{(2)}=\frac{\log\left(\frac{d-2}{d}\right)}{\log\left(\frac{d-1-Q}{d-1+Q}\right)}\, ,
  \label{eq:eqalc2}
\end{equation}
and thus $f^\prime(1)<0$, for all $\alpha <  \alpha_{crit}^{(2)}$, stated differently the solution all-$A$ is stable provided $\alpha <  \alpha_{crit}^{(2)}$.

Let us show that $\alpha_{crit}^{(2)} > \alpha_{crit}^{(1)}$. To prove this claim, we first observe that $\alpha_{crit}^{(1)}$ and $\alpha_{crit}^{(2)}$ share the same numerator. The comparison thus reduces to an analysis of the logarithmic terms appearing in their denominators. In particular, we compute
\begin{eqnarray*}
\frac{d-1-Q}{d-1 + Q}-\frac{(d-1)Q-1}{(d-1)Q + 1}&=&
\frac{2(d-1)(1-Q^2)}{(d-1 + Q)[(d-1)Q + 1]}>0\, ,
\end{eqnarray*}
which is strictly positive since $Q<1$, because of $Q_B<Q_A$.  Then,
\begin{equation*}
 0>\log \left( \frac{d-1-Q}{d-1 + Q} \right) > \log \left (\frac{(d-1)Q-1}{(d-1)Q + 1} \right)\, .
\end{equation*}
Taking the ratio of the corresponding logarithms therefore yields
\begin{equation*}
 \frac{\alpha_{crit}^{(2)}}{ \alpha_{crit}^{(1)}} = \frac{\log \left (\frac{(d-1)Q-1}{(d-1)Q + 1} \right)}{\log \left(\frac{d-1-Q}{d-1+Q} \right)}>1\, ,
\end{equation*}
which proves that $\alpha_{\rm crit}^{(2)} > \alpha_{\rm crit}^{(1)}$.

\begin{figure}[htp!]
    \centering
    \begin{tabular}{c}
    \includegraphics[width=0.42\textwidth]{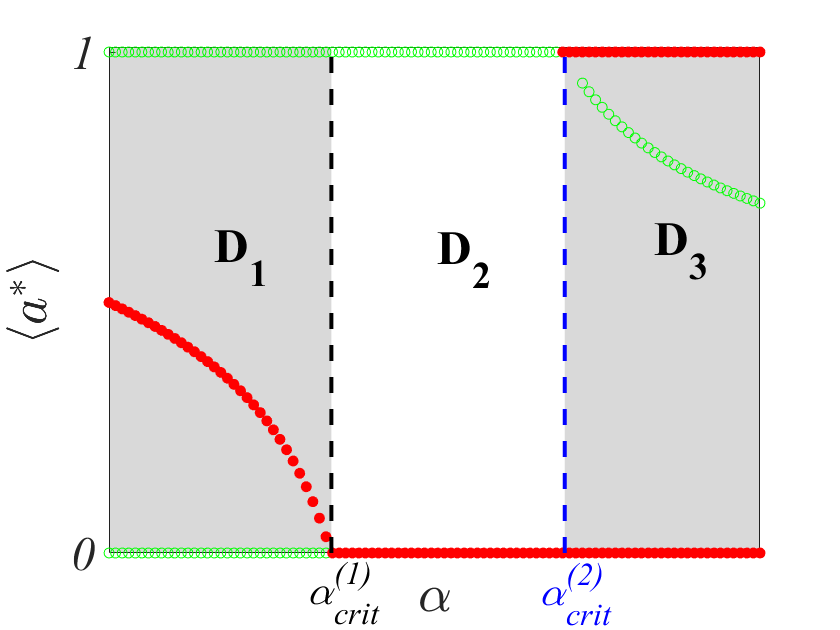} 
    \end{tabular}
        \caption{Recap of the system equilibria and their stability as a function of $\alpha$, highlighting the critical points $\alpha_{\rm crit}^{(1)}$ and $\alpha_{\rm crit}^{(2)}$. Green dots with open markers and red filled dots show the stable and unstable equilibria, respectively. The domains $\mathbf{D_1}$, $\mathbf{D_2}$, and $\mathbf{D_3}$ correspond to the three distinct stability regimes. In $\mathbf{D_1}$, the equilibria $\langle a^* \rangle = 0$ and $\langle a^* \rangle = 1$ are stable, while the non-trivial equilibrium $0 < \langle a^* \rangle < 1$ is unstable. In $\mathbf{D_2}$, the non-trivial equilibrium does not exist and the system admits a single stable equilibrium at $\langle a^* \rangle = 1$, with $\langle a^* \rangle = 0$ being unstable. In $\mathbf{D_3}$, the non-trivial equilibrium $0 < \langle a^* \rangle < 1$ is the only stable one, while the trivial equilibria $\langle a^* \rangle = 0$ and $\langle a^* \rangle = 1$ are unstable. Other parameters are : $d=3$, $Q_A = 1$, and $Q_B=0.75$.}
        \label{fig:figde}
\end{figure}

To summarize the analysis of the stability of the system's equilibria as a function of the control parameter $\alpha$, we identify three distinct dynamical regimes, as illustrated in Fig.~\ref{fig:figde}, where the widths of the different domains may vary depending on the hyperedge size and the quality ratio. Stable (green dots with open markers) and unstable (red filled dots) fixed points are shown, while the critical values $\alpha_{\mathrm{crit}}^{(1)}$ and $\alpha_{\mathrm{crit}}^{(2)}$ are marked by black and blue dashed lines, respectively. \\
\text{(i)} For $\alpha < \alpha_{\rm crit}^{(1)}$ (region $\mathbf{D_1}$), the system exhibits bistability between the two consensus states: both trivial equilibria $\langle a^* \rangle = 0$ and $\langle a^* \rangle = 1$ are stable, while the nontrivial fixed point $0 < \langle a^* \rangle < 1$ is unstable. \\
\text{(ii)} In the intermediate region $\alpha_{\rm crit}^{(1)} < \alpha < \alpha_{\rm crit}^{(2)}$ (region $\mathbf{D_2}$), the system admits a single stable state at $\langle a^* \rangle = 1$, whereas $\langle a^* \rangle = 0$ becomes unstable. This indicates robust consensus formation on option $A$, independent of the initial conditions (as long as at least one agent initially holds opinion $A$).  \\
\text{(iii)} Finally, for $\alpha > \alpha_{\rm crit}^{(2)}$ (region $\mathbf{D_3}$), the nontrivial equilibrium $0 < \langle a^* \rangle < 1$ emerges as the unique stable solution, while both trivial equilibria are unstable. As a consequence, the system no longer supports full consensus and instead settles into a decision deadlock characterized by the coexistence of both opinions.

\begin{figure*}[htp!]
    \centering
    \begin{tabular}{c}
    \includegraphics[width=1.05\textwidth]{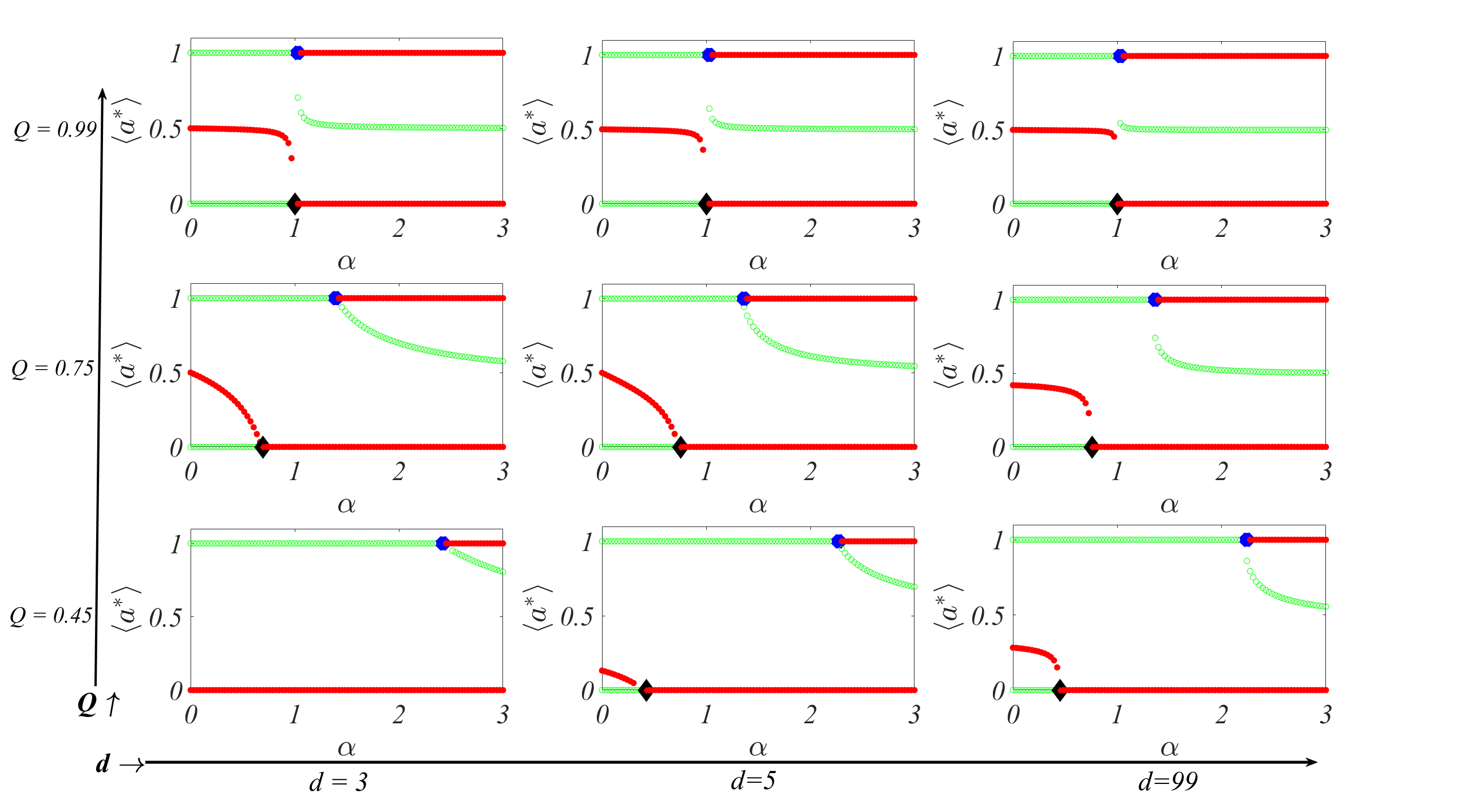} 
    \end{tabular}
        \caption{Equilibria and their stability as a functions of $\alpha$ for different values of $d$ and $Q$: from left to right, the hyperedge size $d$ increases; from bottom to top, the quality ratio $Q$ increases. Green dots with open markers and red filled dots show the stable and unstable equilibria respectively. Black diamonds illustrate the critical value $\alpha_{crit}^{(1)}$ given in Eq.~\ref{eq:eqalc1}, and blue stars indicate the critical value $\alpha_{crit}^{(2)}$ given by Eq.~\ref{eq:eqalc2}.}
        \label{fig:figstabilityalq}
\end{figure*}

A more general analysis is given in Fig.~\ref{fig:figstabilityalq} showing the bifurcation diagram of the analytical model  defined in Eq.~\ref{eq:mainedod}, as a function of the pooling error $\alpha$, for different hyperedge size $d$ and the quality ratio $Q$. From left to right, the hyperedge size $d$ increases; and from bottom to top, the quality ratio $Q$ increases, as indicated by the horizontal (for $d$) and vertical (for $Q$) arrows. The diagram illustrates how the stability of the equilibria changes with $\alpha$ under different structural (hyperedge size) and quality ratio (Q) conditions. In each panel we consider the couple ($d,Q$), remember that once $(d-1)Q>1$, then the equilibrium $\langle a^* \rangle=0$, (i.e., the population always converges to option $B$) is stable (green dots with open markers curve) provided $\alpha < \alpha_{crit}^{(1)}$ (black diamond) and unstable (red filled-dot curve) otherwise; the equilibrium $\langle a^* \rangle=1$ (i.e., the population always converges to option $A$) is stable if $\alpha < \alpha_{crit}^{(2)}$ (denoted by a blue star). In the panel corresponding to ($d=3,Q=0.45$), the condition $(d-1)Q>1$ is not satisfied and thus the equilibrium $\langle a^* \rangle=0$ is never stable; the equilibrium $\langle a^* \rangle=1$ changes its stability at $\alpha=\alpha_{crit}^{(2)}$ (blue star). For all the plots, but for the case just mentioned, the condition $(d - 1)Q > 1$ is always satisfied, and the equilibrium $\langle a^* \rangle = 0$ is stable whenever $\alpha < \alpha_{crit}^{(1)}$.

\begin{figure*}[htp!]
    \centering
    \begin{tabular}{cc}
        \includegraphics[width=0.5\textwidth]{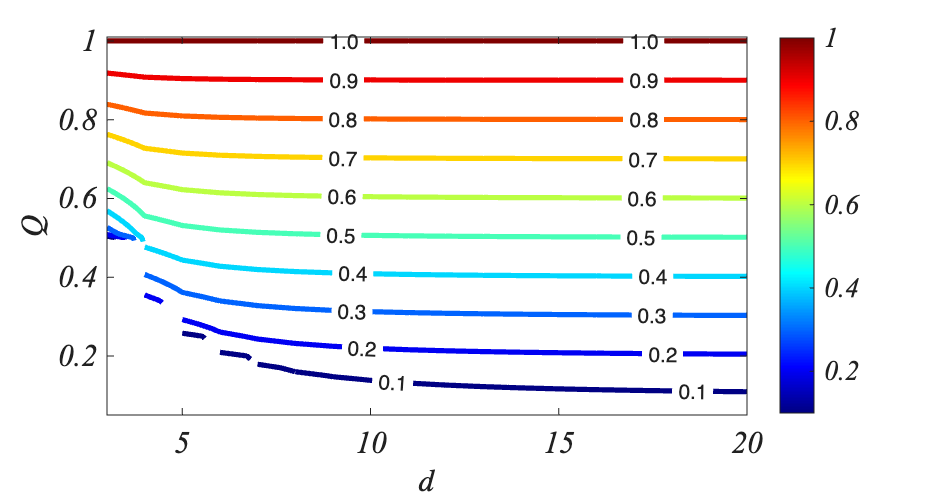} &
        \includegraphics[width=0.5\textwidth]{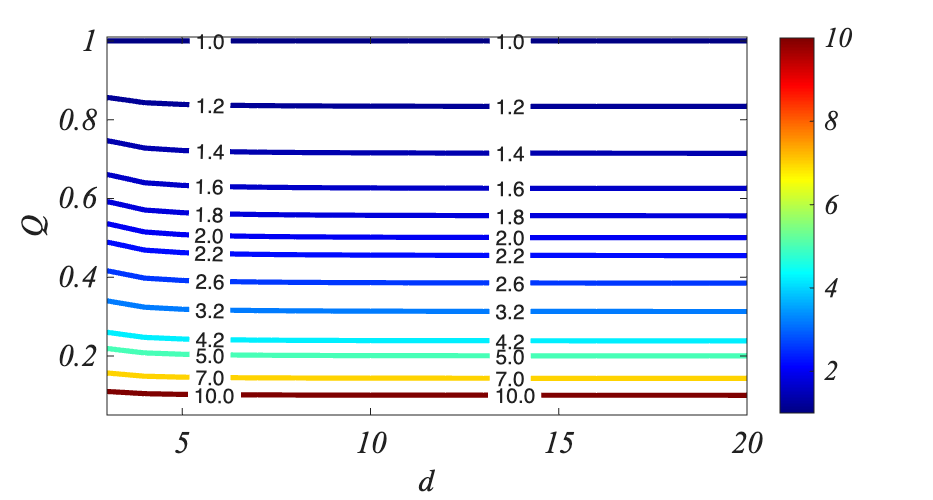}\\
        (a) & (b)
    \end{tabular}
        \caption{Evolution of the critical values: (a) $\alpha^{(1)}_{\text{crit}}$ and (b) $\alpha^{(2)}_{\text{crit}}$ as functions of the parameters $d$ and $Q$. For each pair $(d, Q) \in ([3, 20], [0.1, 1])$, the critical values $\alpha^{(1)}_{\text{crit}}$ (see Eq.~\eqref{eq:eqalc1}) and $\alpha^{(2)}_{\text{crit}}$ (see Eq.~\eqref{eq:eqalc2}) correspond to the thresholds below which the \textit{all-B} (resp. \textit{all-A}) solution is stable. These critical values are shown both on the color bar and along the curves.}
        \label{fig:alpha_crit}
\end{figure*}

From the row corresponding to $Q = 0.45$, we observe that by increasing the hyperedge size can drive the system to converge toward the worst option (i.e., option B); indeed the equilibrium $\langle a^* \rangle=0$ is stable for a slightly larger interval of $\alpha$ values (green branch associated to $\langle a^* \rangle=0$) as $d$ increases. However, this outcome only requires a relatively small initial fraction of agents choosing option $A$. Let us observe that by increasing $d$, the system converges to a deadlock state with a coexistence of agents supporting the better option, $A$, but also the worst one, $B$, for large pooling error $\alpha$. This phenomenon is even more apparent in Fig.~\ref{fig:alpha_crit}, which shows the evolution of the critical values of $\alpha$ associated with the equilibria $\langle a^* \rangle = 0$ (see Fig.~\ref{fig:alpha_crit}(a)) and $\langle a^* \rangle = 1$ (see Fig.~\ref{fig:alpha_crit}(b)). Panel~\ref{fig:alpha_crit}(a) reveals that when the quality ratio $Q < 0.5$, the critical threshold $\alpha_{\mathrm{crit}}^{(1)}$ does not exist for small group sizes. In this case, the system converges to the equilibrium $\langle a^* \rangle = 1$ (i.e., all agents choose $A$), even if the initial proportion of agents choosing $A$ is vanishingly small. This behavior can be observed in the particular case of $(Q = 0.45,\, d = 3)$ shown in Fig.~\ref{fig:figstabilityalq}, where the  equilibrium $\langle a^* \rangle = 0$ is always unstable (red filled-dot curve).
However, the hyperedge size does not significantly  affect the existence of the equilibrium $\langle a^* \rangle = 1$, but rather shifts minimally the critical value at which it appears, as shown in Fig.~\ref{fig:figstabilityalq}(b). This figure illustrates the strong influence of the quality ratio $Q$ on the critical threshold $\alpha_{\mathrm{crit}}^{(2)}$. These results highlight not only the effect of hyperedge size on the system’s outcome but also elucidate the role of the quality ratio $Q$, which measures the difficulty of the decision problem: the larger $Q$ is, the harder it becomes for the group to vote the best option. As $Q \to 1$, the decision problem becomes increasingly difficult, and the critical values $\alpha_{\mathrm{crit}}^{(1)}$ and $\alpha_{\mathrm{crit}}^{(2)}$ converge towards the same value, $\alpha_{\mathrm{crit}}^{(1)} \approx \alpha_{\mathrm{crit}}^{(2)} \approx 1$, independently of the  hyperedge size within the network (see both panels in Fig.~\ref{fig:alpha_crit}). This increasing complexity as $Q \to 1$ has a dual consequence on collective decision-making: depending on the initial conditions $a_{\mathrm{ini}}$, the system may converge either to the best option $\langle a^* \rangle = 1$, i.e., option A, or to the worst option $\langle a^* \rangle = 0$, i.e., option B (for instance, for $a_{\mathrm{ini}} = 0.5$ and $\alpha < \alpha_{\mathrm{crit}}^{(1)} \approx \alpha_{\mathrm{crit}}^{(2)}$), or it may evolve towards a coexistence state of both opinions (in this case, regardless of the initial condition $a_{\mathrm{ini}}$, for $\alpha > \alpha_{\mathrm{crit}}^{(1)} \approx \alpha_{\mathrm{crit}}^{(2)}$), as shown in the line $Q = 0.99$ in Fig.~\ref{fig:figstabilityalq}.

\subsubsection{Agent Based Model on $d$-uniform hypergraph} \label{sec:ABMH}

To corroborate the results obtained from the analytical theory under the assumption of HMF for hypergraphs, we perform numerical simulations based on an agent-based model. The system consists of a population of $N=1000$ agents interacting in groups, represented by a $d$-uniform hypergraph with $1500$ hyperedges for both random and scale-free topologies, the latter being generated with a power-law exponent $\gamma = 2.5$ (see Appendices~\ref{sec: appps1}--\ref{sec: appps2} for the construction algorithms of random and scale-free $d$-uniform hypergraphs). Each agent holds an opinion, either $A$ or $B$, with an associated quality $Q_A$ or $Q_B$, respectively. Opinions are initially assigned randomly to the nodes of the network, which consequently leads to a random distribution of opinions within the hyperedges. The opinion of a randomly selected agent is updated according to the probability $P_{\alpha}$ defined in Eq.~\ref{eq:Px}. Each numerical simulation (experiment) runs for a total time of $T_{\text{max}} = 10^5$ iterations and allows to compute a total weighted fraction $\rho_A$ (resp. $\rho_B$) of agents holding opinion $A$ (resp. $B$), namely the analogous of the quantity $\langle a\rangle$ above defined, given by
\begin{equation}
\begin{split}
   \rho_A(T_{\text{max}})   = \sum_{k \geq 0} q_k \frac{n^{k+1}_A(T_{\text{max}})}{k+1} \\
    \text{(resp. } \rho_B(T_{\text{max}}) = 1 - \rho_A(T_{\text{max}})).
\end{split}
\label{eq:rhop}
\end{equation}
where $n^{k+1}_A(T_\text{max})$ is the number of agents with hyperdegree $k+1$ and opinion $A$ at time $T_\text{max}$, and $q_k$ is the excess hyperdegree expressed in Eq.~\ref{eq:excess}. 
The results are repeated $n = 50$ times, independently each other; the obtained value of $\rho^{(j)}_A(T_{\text{max}})$ at the $j$-th simulation, are used to compute the average value $\bar{\rho}_A(T_{\text{max}})$ (respectively $\bar{\rho}_B(T_{\text{max}})$) given by
\begin{equation}
\begin{split}
    \bar{\rho}_A(T_{\text{max}}) &= \frac{1}{n} \sum_{j=1}^n \rho^j_A(T_{\text{max}}) \\
    \text{(resp. } \bar{\rho}_B(T_{\text{max}}) &= \frac{1}{n} \sum_{j=1}^n \rho^j_B(T_{\text{max}}) \text{)}
\end{split}
\label{eq:mu}
\end{equation}

\begin{figure*}[htp!]
\centering
\begin{tikzpicture}
\node[anchor=south west, inner sep=0] (tab) at (0,0) {
\begin{tabular}{ccc}
\includegraphics[width=0.3\textwidth]{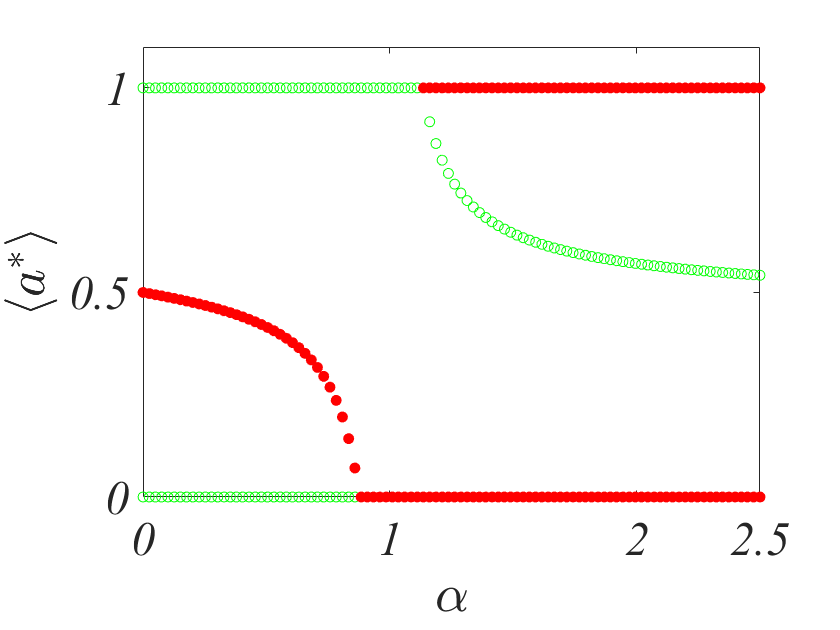} &
\includegraphics[width=0.3\textwidth]{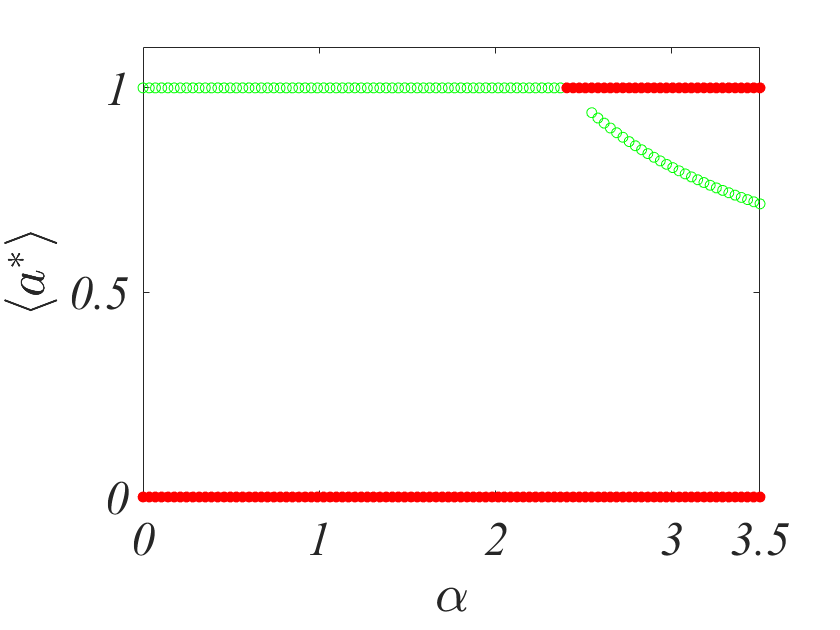} &
\includegraphics[width=0.3\textwidth]{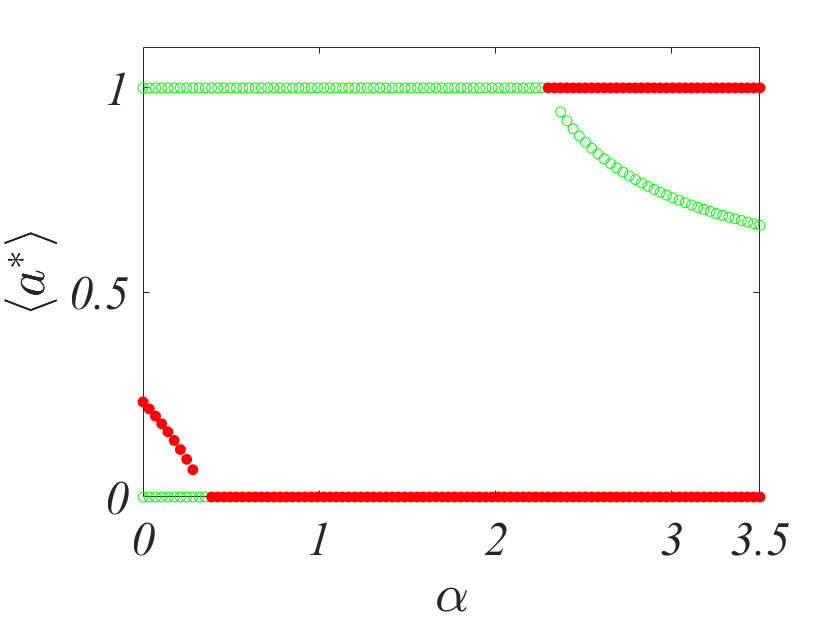} \\
\hspace{4 mm}(a) & (b) & (c) \\
\includegraphics[width=0.3\textwidth]{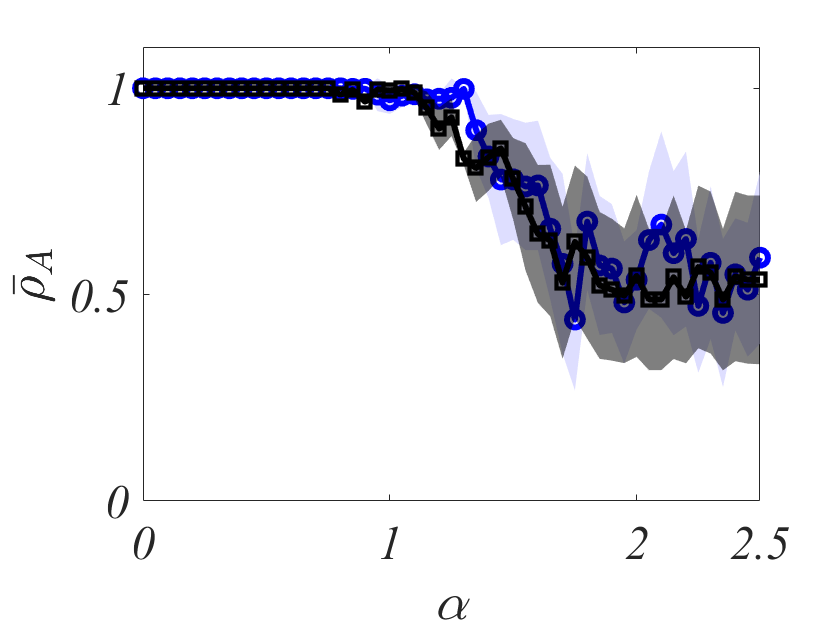} &
\includegraphics[width=0.3\textwidth]{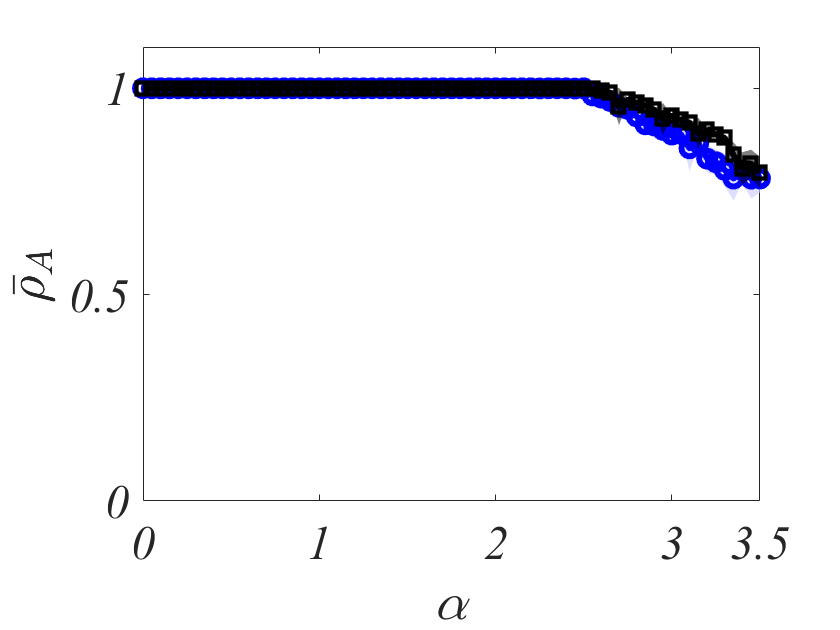} &
\includegraphics[width=0.3\textwidth]{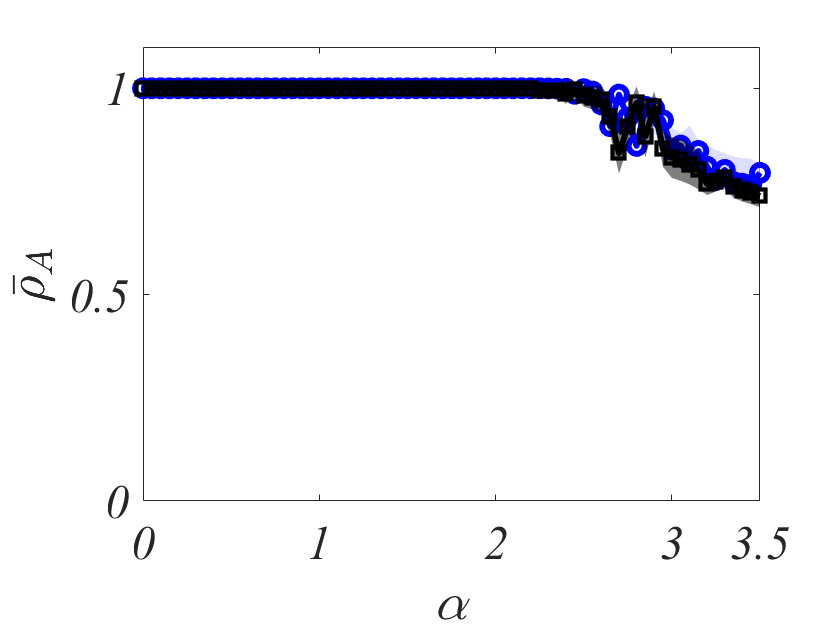} \\
(d) & (e) & (f)
\end{tabular}
};
\begin{scope}[x={(tab.south east)}, y={(tab.north west)}]
\draw[magenta, ultra thick, dashed] (0.166,0.15) -- (0.166,0.97);
\draw[magenta, ultra thick, dashed] (0.558,0.15) -- (0.558,0.97);
\draw[magenta, ultra thick, dashed] (0.88,0.15) -- (0.88,0.97);
\node[anchor=south, yshift = -20 pt] at (0.166,1.05) {$(d = 3, Q_B = 0.9)$};
\node[anchor=south, yshift = -20 pt] at (0.5,1.05)   {$(d = 3, Q_B = 0.45)$};
\node[anchor=south, yshift = -20 pt] at (0.833,1.05) {$(d = 4, Q_B = 0.45)$};
\node at (0.5,-0.05) {
\begin{tikzpicture}[baseline]
    \draw[green] (0,0) circle (2pt);
    \node[right] at (0.01,0) {Stable equilibria};

    \draw[red, fill=red] (0.2,0) circle (2pt);
    \node[right] at (0.205,0) {Unstable equilibria};

    \draw[blue, line width=1pt] (0.425,0) -- (0.455,0);
    \draw[blue] (0.44,0) circle (2pt);
    \node[right] at (0.465,0) {Random hypergraphs};

    \draw[black, line width=1pt] (0.70,0) -- (0.73,0);
    \draw[black, very thick, fill=none] (336.715-2pt, 0-2pt) rectangle (336.715+2pt,0+2pt);
    \node[right] at (0.735,0) {Scale-free hypergraphs};
\end{tikzpicture}
};
\end{scope}
\end{tikzpicture}
\caption{We present the bifurcation diagram and the average fraction of agents holding opinion $A$ as a function of the parameter $\alpha$, for both $d$-uniform random and scale-free hypergraphs. The first row (panels a–c) displays the results of the analytical model, while the second row (panels d–f) shows results from the agent-based model (ABM): blue dotted lines with open markers correspond to random $d$-uniform hypergraphs, and black lines with square markers correspond to scale-free $d$-uniform hypergraphs, with $N = 1000$ nodes. The associated standard deviations are shown as transparent blue and black shaded regions, respectively. From left to right, we consider the following parameter settings: $d = 3$ and $Q_B = 0.9$ (panels a, d), $d = 3$ and $Q_B = 0.45$ (panels b, e), and $d = 4$ and $Q_B = 0.45$ (panels c, f). Shaded regions represent the standard deviation. Stable and unstable equilibria are represented by green dots with open markers and red filled dots, respectively in the first row. The quality of opinion $A$ is fixed at $Q_A = 1$. The magenta vertical line indicates the theoretical critical value $\alpha_{\mathrm{crit}}^{(2)}$, thus enabling a direct comparison with the ABM results. In the ABM model, the results are averaged over $50$ independent simulations, each run for a maximum simulation time of $T_{\text{max}} = 10^5$ iterations.}
\label{fig::f3}
\end{figure*}

Fig.~\ref{fig::f3} provides a comparative analysis between $\langle a^*\rangle$ for the analytical model (top row, panels a–c) and $\bar{\rho}_A(T_{\text{max}})$ for the agent-based simulations (bottom row, panels d–f) as a function of the pooling error, $\alpha$, across different configurations of the hyperedge size $d$ and the quality ratio $Q$. Let us recall that $\alpha$ measures the degree of error with which agents assess the overall quality of an opinion based on the opinions present in their group. A low value of $\alpha$ indicates an accurate perception of collective quality, while a high value reflects a misjudgment, introducing noise or social misinterpretation. From left to right, the panels correspond to three configurations: $(d = 3, Q = 0.9)$ in panels~\ref{fig::f3}(a, d), $(d = 3, Q = 0.45)$ in panels~\ref{fig::f3}(b, e), and $(d = 4, Q = 0.45)$ in panels~\ref{fig::f3}(c, f). In all simulations, initial conditions are symmetric, with half the agents holding opinion $A$ and half holding $B$, randomly assigned to the nodes. For the ABM results, blue dotted lines with open markers correspond to random $d$-uniform hypergraphs, and black lines with square markers correspond to scale-free hypergraphs. Shaded areas represent standard deviations, averaged over $50$ independent runs with a maximum simulation time of $T_{\text{max}} = 10^5$ iterations. Across all panels, the top-row diagrams exhibit the equilibria analytically derived by using the HMF: green dots with open markers represent stable equilibria, and red filled dots dots denote unstable ones as presented in the previous section. These bifurcation diagrams highlight the critical thresholds $\alpha_{\text{crit}}^{(1)}$ and $\alpha_{\text{crit}}^{(2)}$, beyond which qualitative changes in system behavior occur. Notably, in panel~\ref{fig::f3}(a), where $(d = 3, Q_B = 0.9)$, a bifurcation leads to the emergence of two stable equilibria, zero and one for $\alpha<\alpha_{\text{crit}}^{(1)} \approx 0.88$ as illustrated in Fig.~\ref{fig:alpha_crit}(a), one stable equilibrium for $\alpha_{\text{crit}}^{(1)} \approx 0.88 < \alpha< \alpha_{\text{crit}}^{(2)} \approx 1.2$, and one stable non trivial equilirium for $\alpha > \alpha_{\text{crit}}^{(2)} \approx 1.2$.  This feature is mirrored by the ABM in panel~\ref{fig::f3}(d). In panel~\ref{fig::f3}(b), the condition $(d - 1)Q < 1$ results in the stability of only one branch, and both the analytical and ABM results (panel~\ref{fig::f3}e) confirm that the system converges to the higher-quality opinion for $\alpha<\alpha_{\text{crit}}^{(2)} \approx 2.4$. The same qualitative agreement is observed in panels~\ref{fig::f3}(c) and~\ref{fig::f3}(f) for $d = 4$ and $Q_B = 0.45$, although the increased hyperedge size introduces a smoother transition in the ABM outcomes. Overall, the consistency between the analytical and simulation results is remarkable, reinforcing the validity of the theoretical model in capturing the macroscopic dynamics, even under varying network topologies (random vs. scale-free) and system sizes. Let us observe that panels~\ref{fig::f3}(d) and~\ref{fig::f3}(f) do not show the possible convergence of the system to the equilibrium $\langle a^*\rangle=0$ present in the top left and top right panels for small $\alpha$; this is because to achieve this equilibrium, the system requires a small enough initial population of agents supporting option $A$, and this is not how the ABM has been initialized, i.e., $50\%$ of agents with opinion $A$ and $50\%$ with opinion $B$.

\begin{figure*}[htp!]
\centering
\begin{tikzpicture}
\node[anchor=south west, inner sep=0] (tab) at (0,0) {
\begin{tabular}{cc}
\hspace{-2.10cm}
\includegraphics[width=0.55\textwidth]{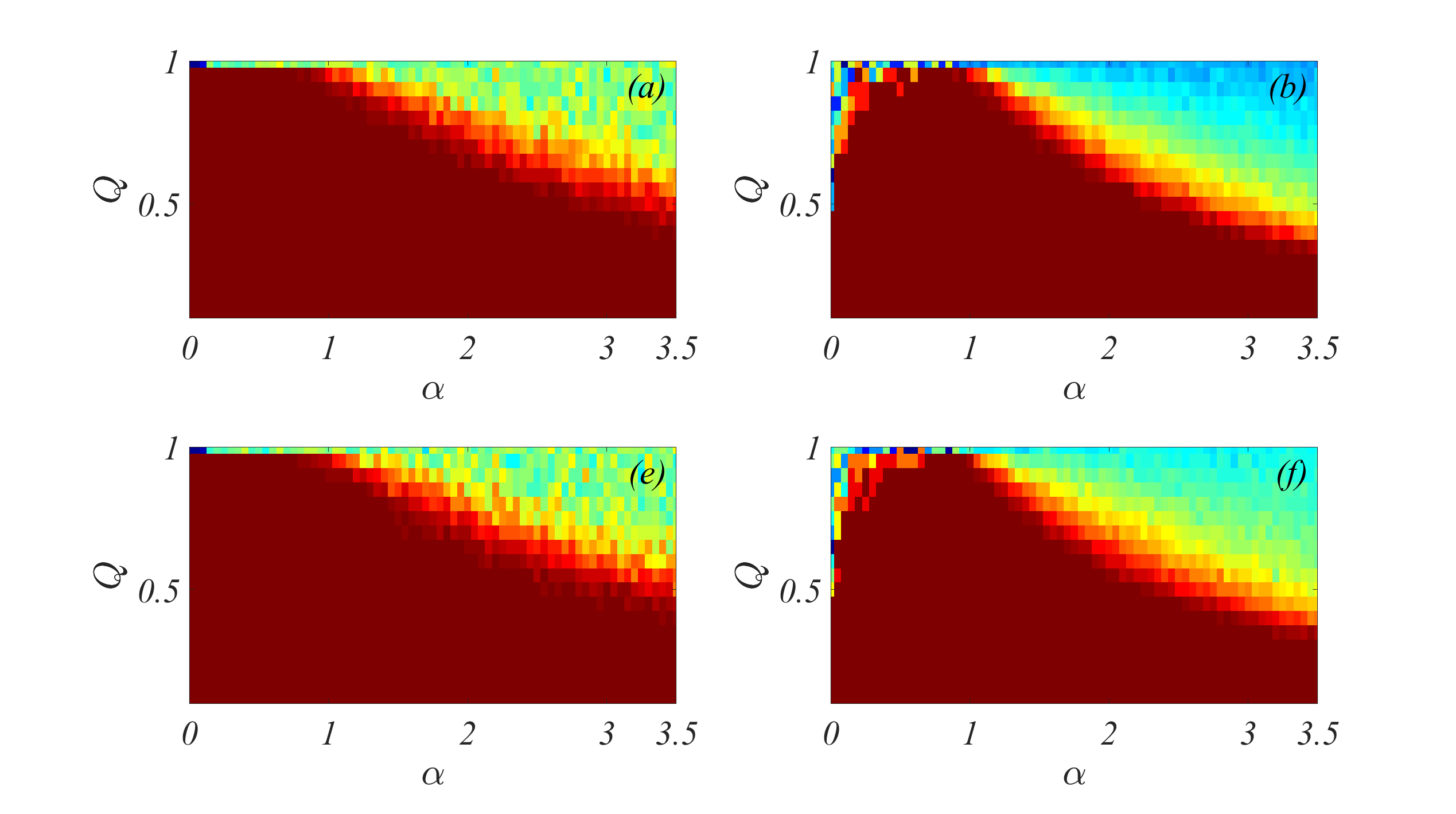} &
\hspace{-1.05cm}
\includegraphics[width=0.55\textwidth]{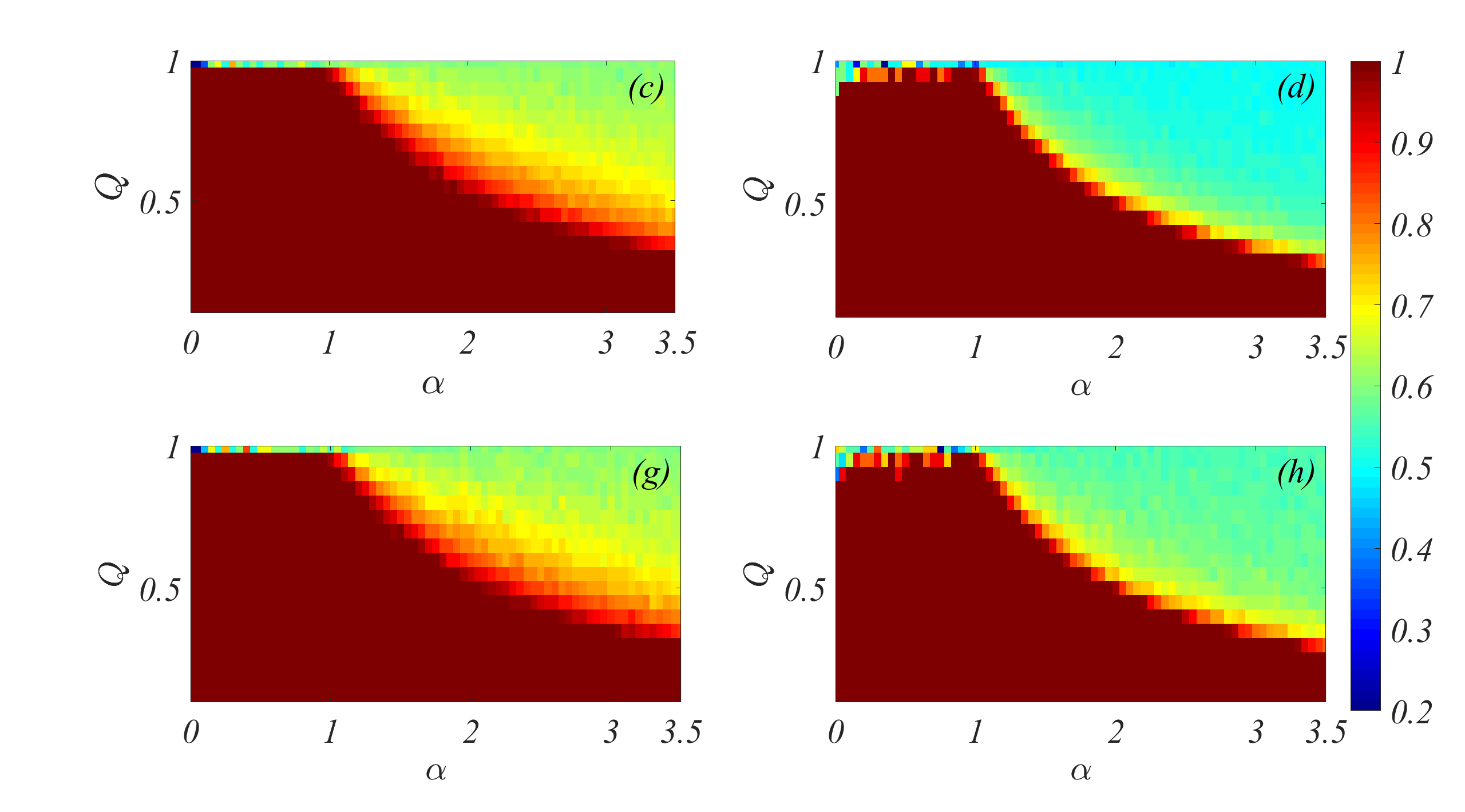}
\end{tabular}
};
\hspace{-1.0cm}
\begin{scope}[x={(tab.south east)}, y={(tab.north west)}]
\draw[-{Stealth[length=6pt,width=6pt]}, line width=1.5pt](0.05,0.045) -- (1.0,0.045);
\draw[-{Stealth[length=6pt,width=6pt]}, line width=1.5pt](-0.02,0.20) -- (-0.02,0.90);
\node[anchor=north] at (0.12,0.03)  {\textit{d=}3};
\node[anchor=north] at (0.374,0.03)  {\textit{d=}4};
\node[anchor=north] at (0.645,0.03)  {\textit{d=}5};
\node[anchor=north] at (0.9,0.03)  {\textit{d=}20};
\node[anchor=east] at (-0.025,0.750) {\textit{RH}};
\node[anchor=east] at (-0.025,0.30) {\textit{SFH}};
\end{scope}
\end{tikzpicture}
\caption{Diagram showing the average fraction of agents holding opinion $A$, $\bar{\rho}_A$, on both $d$-uniform random hypergraphs (panels~(a--d), labelled RH) and $d$-uniform scale-free hypergraphs (panels~(e--h), labelled SFH), as a function of the parameters $\alpha$ and $Q$ (with $Q = Q_B / Q_A$). From left to right, the hyperedge size increases: panels~(a,e) correspond to $d = 3$, panels~(b,f) to $d = 4$, panels~(c,g) to $d = 5$, and panels~(d,h) to $d = 20$. The hypergraph size is fixed at $N = 1000$ for both random and scale-free hypergraphs. Each result is averaged over 50 independent simulations, with a maximum simulation time of $T_{\mathrm{max}} = 10^5$ iterations per experiment. }
\label{fig::ABM2drdsf}
\end{figure*}

To obtain a global view of the system's fate as a function of the main parameters, Fig.~\ref{fig::ABM2drdsf} illustrates the average fraction of agents holding opinion $A$, $\bar{\rho}_A$, obtained from the Agent Based Model on a $d$-uniform random hypergraph (see panels~\ref{fig::ABM2drdsf}(a–d)) and on a scale-free $d$-uniform hypergraph (see panels~\ref{fig::ABM2drdsf}(e–h)), as a function of the pooling error $\alpha$ and the quality ratio $Q$. From left to right, the panels show results for different hyperedge sizes in both network types: 
Panels~\ref{fig::ABM2drdsf}(a,e) correspond to $d=3$, Panels~\ref{fig::ABM2drdsf}(b,f) to $d=4$, 
Panels~\ref{fig::ABM2drdsf}(c,g) to $d=5$, and Panels~\ref{fig::ABM2drdsf}(d,h) to $d=20$. 
Qualitatively, the results are nearly identical across the two types of hypergraphs, confirming the prediction of the analytical model presented above, which indicates that 
the system's behavior is influenced by the hyperedge size rather than the type of hypergraph. The same results also show that, for low pooling error levels ($\alpha \lesssim 1$), most agents tend to converge toward the best option, i.e., the one associated with the higher-quality opinion $A$ (red region). This convergence becomes even more pronounced as $Q$ decreases ($Q \ll 1$), corresponding to an easy decision problem as $Q_A>Q_B$, indeed the region of convergence to all-$A$ (red region) exists for larger value of $\alpha$. Conversely, when $Q$ approaches unity (i.e, $Q_A \approx Q_B$), the decision process becomes increasingly difficult due to the similarity in the quality of the two opinions. This leads to the coexistence of opinions $A$ and $B$ for vanishing pooling error, as illustrated in the panels of Fig.~\ref{fig::ABM2drdsf} by values of $\bar{\rho}_A$ lower than one (see the color scale).
However, as the pooling error $\alpha$ increases, the likelihood for the system to collectively select the best option---option $A$---decreases with increasing $Q$. Even for relatively simple decision problems, for instance $Q \approx 0.7$, larger pooling errors lead the system to converge toward a decision dead-lock state characterized by the coexistence of opinions $A$ and $B$.\\
These results reveal the effect of the parameter $d$, which characterizes the hyperedge (or group) size. As $d$ increases, the yellow–lime green region corresponding to $\bar{\rho}_A \approx [0.45, 0.65]$, representing the coexistence of the two opinions, becomes broader. This result shows that, for the same  pooling error and even for moderate decision difficulty, larger groups exhibit a reduced ability to reach unanimous consensus. This highlights a paradox of large groups: instead of benefiting from collective intelligence, they may become more susceptible to collective error, leading the system into an indecisive state where opinions $A$ and $B$ coexist.

\begin{figure}[htp!]
    \centering
    \begin{tabular}{c}
    \hspace{-0.60cm}
        \includegraphics[width=0.5\textwidth]{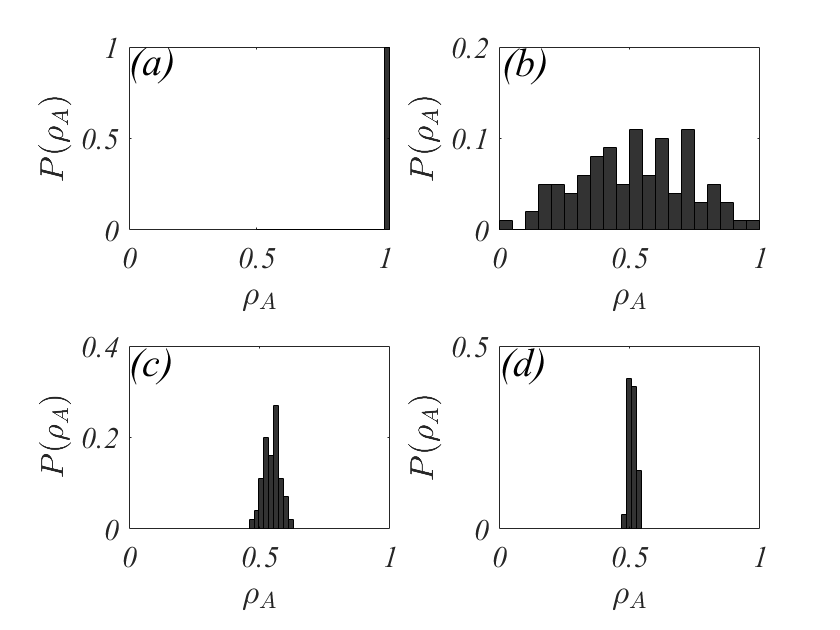} 
    \end{tabular}
        \caption{Distribution of $\rho_A$ from $n = 100$ independent experiments conducted on a random $d$-uniform hypergraph for different hyperedge sizes, $\alpha$, and $Q$.  
        Panel~(a) shows the case $(\alpha = 1.0,\ Q = 0.45)$, where the distribution remains similar for all hyperedge sizes $d \geq 3$.  
        Panels~(b)--(d) correspond respectively to $d = 3$, $d = 5$, and $d = 20$, with $(\alpha = 3.0,\ Q = 0.8)$.}
        \label{fig::HHMFscheme}
\end{figure}

Fig.~\ref{fig::HHMFscheme} displays the distribution of the fraction of agents supporting opinion $A$, obtained from $n = 100$ independent simulations on a random $d$-uniform hypergraph. 
Panel~\ref{fig::HHMFscheme}(a) corresponds to the parameter choice $(\alpha = 1.0,\ Q = 0.45)$, under which the system systematically reaches consensus in favor of opinion $A$, regardless of the group size $d \geq 3$. This is reflected by a sharp peak at $\rho_A = 1$, indicating unanimous adoption of opinion $A$ across all realizations as illustrated in Fig.~\ref{fig::ABM2drdsf}. Panels~\ref{fig::HHMFscheme}(b) through ~\ref{fig::HHMFscheme}(d) show the distributions for a more challenging scenario with $(\alpha = 3.0,\ Q = 0.8)$, respectively for $d = 3$, $5$, and $20$. 
In panel~\ref{fig::HHMFscheme}(b), the distribution is close to a bimodal, meaning that across the replicas the system sometimes converged to the all-$A$ case and sometime else to the all-$B$ one, observe also that in several cases the system went to a deadlock state. As the group size increases, the distributions become unimodal and increasingly centered around $\rho_A = 0.5$ (panels~\ref{fig::HHMFscheme}(c) and ~\ref{fig::HHMFscheme}(d)), reflecting a stable coexistence of both opinions. 

Collective decision-making is characterized by the emergence of consensus, defined as the state in which all agents within a group share the same opinion or select the same option, thus marking the end of the decision process. The convergence time corresponds to the duration required for the system to reach this consensus, that is, the moment when opinions stop changing. Computing this time allows for the evaluation of the speed of the collective decision-making process. This measure is therefore essential to understand and optimize the performance of natural or artificial collective systems, as it directly links the group agreement speed (or decision time) with both the pooling error, $\alpha$ and the difficulty of the decision problem (quality ratio, $Q$).

\begin{figure*}[htp!]
\centering
\begin{tikzpicture}
\node[anchor=south west, inner sep=0] (tab) at (0,0) {
\begin{tabular}{cc}
\hspace{-2.10cm}
\includegraphics[width=0.55\textwidth]{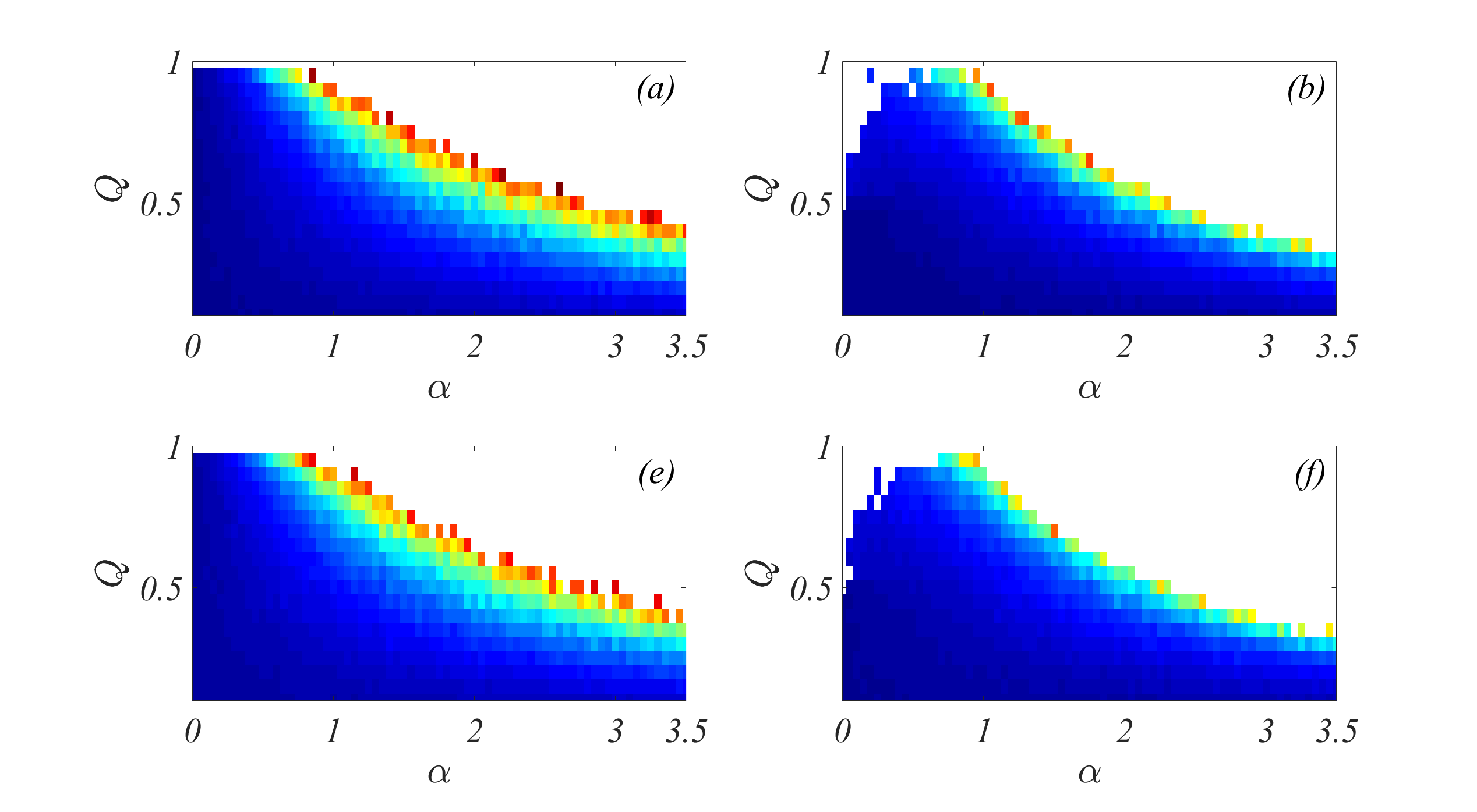} &
\hspace{-1.05cm}
\includegraphics[width=0.55\textwidth]{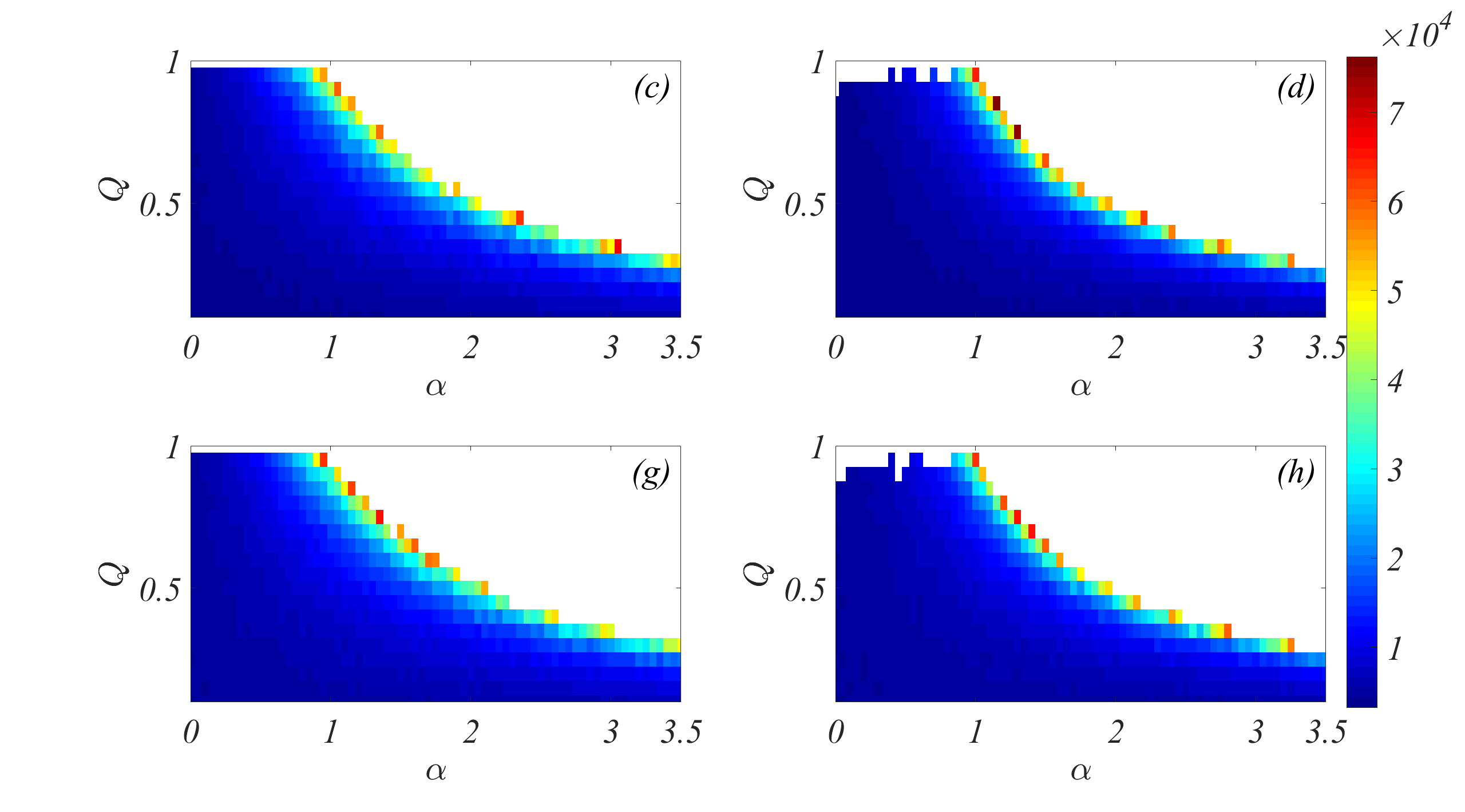}
\end{tabular}
};
\hspace{-1.0cm}
\begin{scope}[x={(tab.south east)}, y={(tab.north west)}]
\draw[-{Stealth[length=6pt,width=6pt]}, line width=1.5pt](0.05,0.045) -- (1.0,0.045);
\draw[-{Stealth[length=6pt,width=6pt]}, line width=1.5pt](-0.02,0.10) -- (-0.02,0.90);
\node[anchor=north] at (0.12,0.03)  {\textit{d=}3};
\node[anchor=north] at (0.374,0.03)  {\textit{d=}4};
\node[anchor=north] at (0.645,0.03)  {\textit{d=}5};
\node[anchor=north] at (0.9,0.03)  {\textit{d=}20};
\node[anchor=east] at (-0.025,0.750) {\textit{RH}};
\node[anchor=east] at (-0.025,0.30) {\textit{SFH}};
\end{scope}
\end{tikzpicture}
\caption{Decision time on both $d$-uniform random hypergraphs (panels~(a--d), labelled RH) and $d$-uniform scale-free hypergraphs (panels~(e--h), labelled SFH), as a function of the parameters $\alpha$ and $Q$ (with $Q = Q_B / Q_A$). From left to right, the hyperedge size increases: panels~(a,e) correspond to $d = 3$, panels~(b,f) to $d = 4$, panels~(c,g) to $d = 5$, and panels~(d,h) to $d = 20$. The hypergraph size is fixed at $N = 1000$ for both random and scale-free hypergraphs. Each result is averaged over 50 independent simulations, with a maximum simulation time of $T_{\mathrm{max}} = 10^5$ iterations per experiment.}
\label{fig::ABM2drdsftime}
\end{figure*}
Fig.~\ref{fig::ABM2drdsftime} shows the average time required to reach consensus on option $A$ in the agent-based model 
as a function of the pooling error $\alpha$ and task difficulty $Q$, corresponding to the results 
presented in Fig.~\ref{fig::ABM2drdsf}. From left to right, the panels illustrate different hyperedge sizes 
for both network types: Panels~\ref{fig::ABM2drdsftime}(a,e) correspond to $d=3$, 
Panels~\ref{fig::ABM2drdsftime}(b,f) to $d=4$, Panels~\ref{fig::ABM2drdsftime}(c,g) to $d=5$, 
and Panels~\ref{fig::ABM2drdsftime}(d,h) to $d=20$.
The white regions indicate parameter settings where no consensus is reached in the allowed time for the simulation, this implies that opinions $A$ and $B$ coexist (see Fig.~\ref{fig::ABM2drdsf}).
These findings indicate that collective decision-making, i.e., consensus is reached more rapidly for low pooling errors (lower $\alpha$) and for easier tasks (lower $Q$), a behavior commonly observed in collective decision-making dynamics where stronger effective signals facilitate faster 
agreement~\cite{couzin2011uninformed} with the network. Furthermore, from left to right, the consensus time decreases with increasing hyperedge size $d$, highlighting the role of group interactions in accelerating opinion convergence, as reported in previous studies on higher-order social interactions and group decision processes~\cite{battiston2020networks}. Notably, the behavior is qualitatively similar across random and scale-free $d$-uniform hypergraphs, confirming that the dynamics are largely determined by hyperedge size and task parameters rather 
than the specific network topology, consistent with our claim.

\subsection{ABM on heterogeneous hypergraph} \label{sec:ABMGH}

In this section, we investigate the ABM on \textit{heterogeneous hypergraphs}, where nodes can participate in hyperedges of varying sizes, unlike the $d$-uniform case studied so far, where hyperedge sizes were fixed at $d$. These heterogeneous structures capture more diverse higher-order interactions. Given the difficulty of constructing an analytical mathematical model that could serve as a predictor for such heterogeneous hypergraphs, this study is entirely numerical. We analyze how structural heterogeneity influences collective decision-making dynamics. For this analysis, we consider a hypergraph of size $N=1000$ in which the hyperedge sizes, i.e., \textit{the number of nodes $+1$} per hyperedge, uniformly vary between $d_{\min}$ and $d_{\max}$. In this specific case, we set $d_{\min}=3$ and $d_{\max}=10$, although this can be extended to any hyperedge size range.

\begin{figure*}[htp!]
    \centering
    \begin{tabular}{cc}
    \hspace{-0.80cm}
    \includegraphics[width=0.55\textwidth]{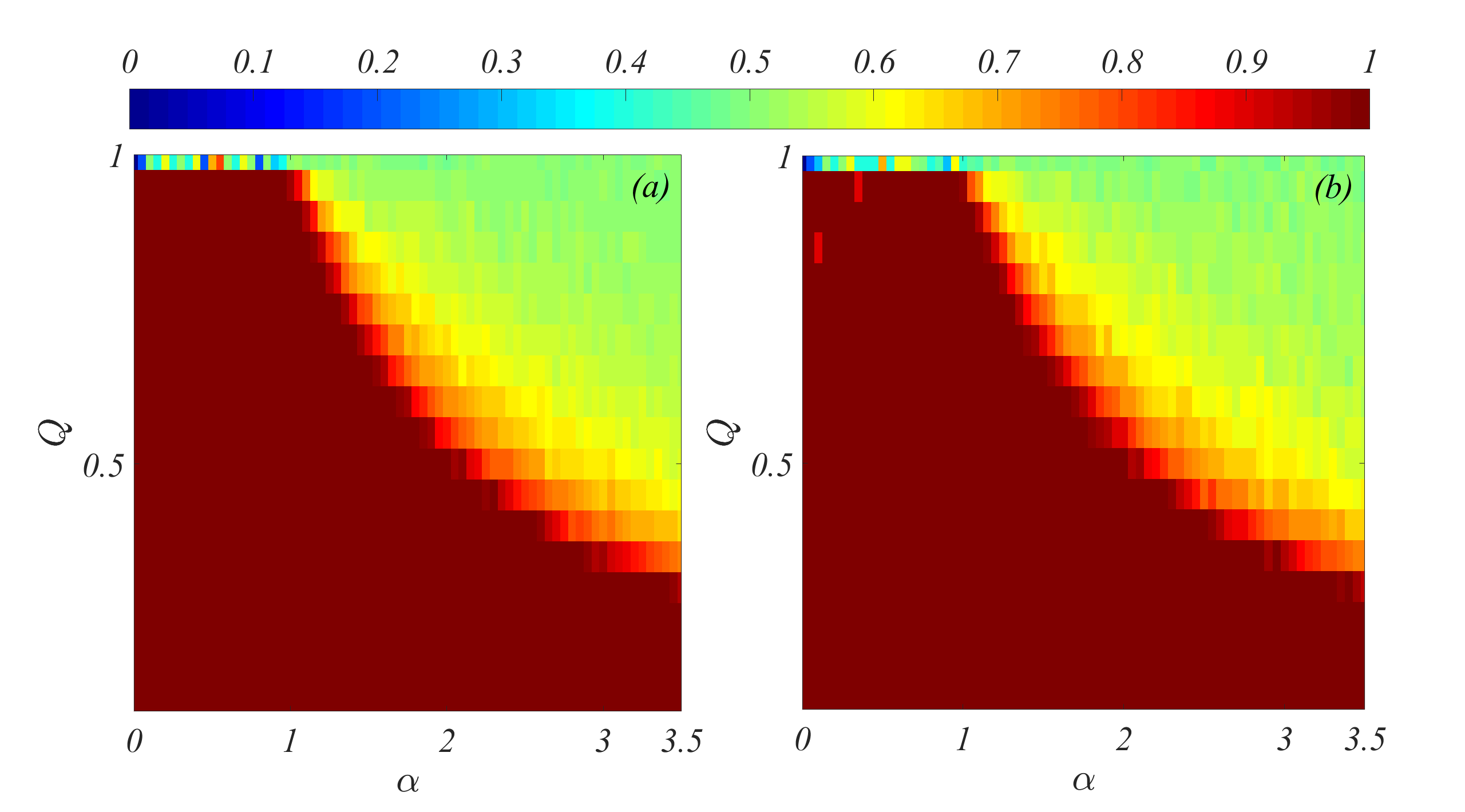}&
    \hspace{-0.80cm}
    \includegraphics[width=0.56\textwidth]{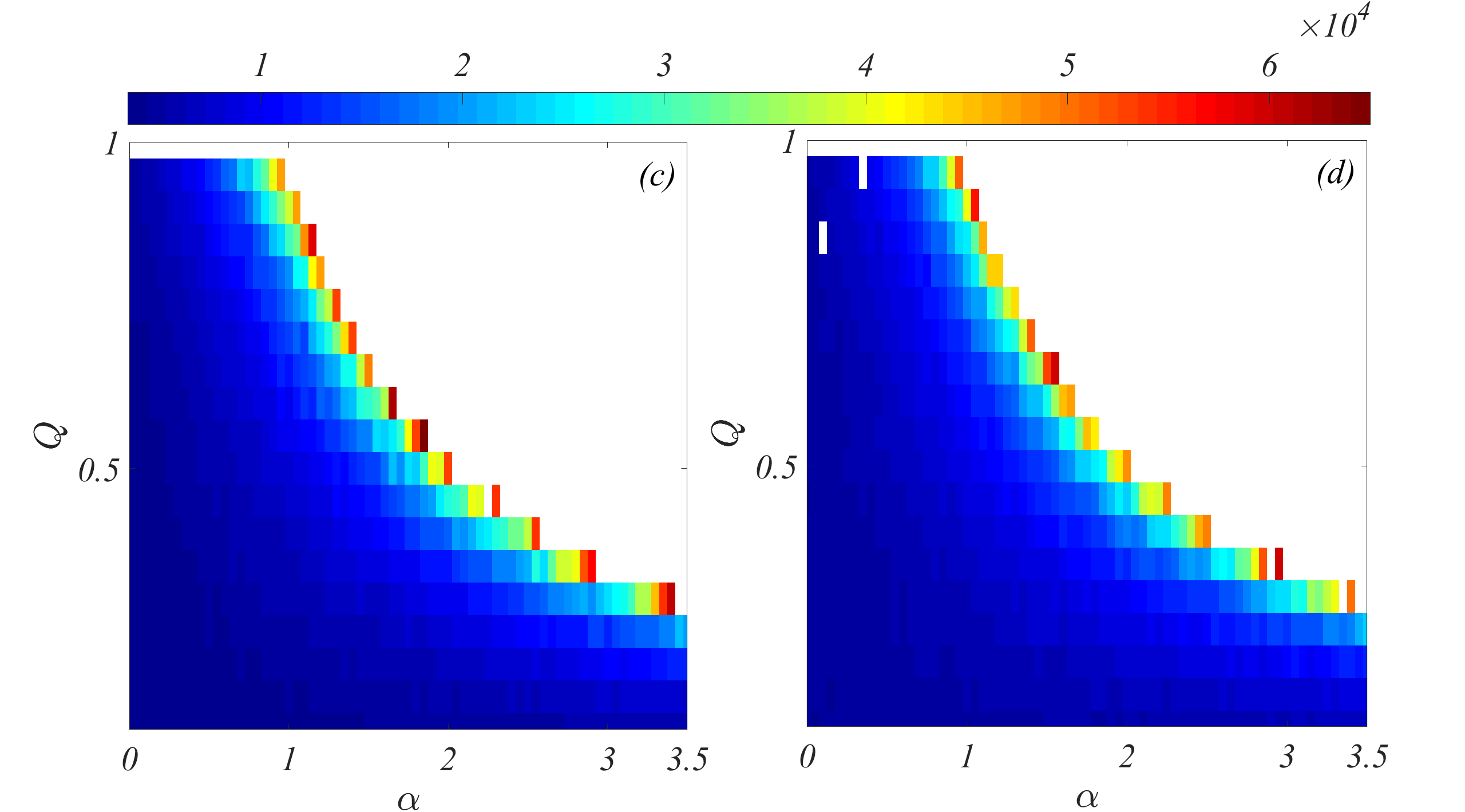}
    \end{tabular}
    \caption{Illustration of the average fraction of agents holding opinion $A$, $\bar{\rho}_A$, and the decision time on heterogeneous hypergraphs as a function of the parameters $\alpha$ and $Q$ (with $Q = Q_B / Q_A$). Panels (a) and (c) correspond to random hypergraphs, while panels (b) and (d) correspond to scale-free hypergraphs. Hyperedge sizes are uniformly drawn from the interval $[d_{\min}, d_{\max}] = [3, 10]$. Each data point represents an average over 30 independent simulation runs, with a maximum simulation time of $T_{\text{max}} = 5 \times 10^5$ iterations per experiment.}
    \label{fig::ABMhetero_a}
\end{figure*}

A systematic study based on an ABM was conducted to analyze the collective decision-making dynamics on higher-order and heterogeneous interaction structures.
Fig.~\ref{fig::ABMhetero_a} summarizes the collective outcome of the agent-based dynamics as a function of the pooling error $\alpha$ and the quality ratio $Q$. 
Panels~\ref{fig::ABMhetero_a}(a) and ~\ref{fig::ABMhetero_a}(b) report the average fraction of agents holding opinion $A$, $\bar{\rho}_A$ (see colorbar), for random and scale-free heterogeneous hypergraphs respectively, while panels~\ref{fig::ABMhetero_a}(c) and ~\ref{fig::ABMhetero_a}(d) display the corresponding average decision time (see colorbar). Results are averaged over $50$ independent realizations, with simulations run up to $T_{\text{max}} = 5 \times 10^5$ to ensure convergence to the asymptotic regime. For sufficiently small pooling errors and  quality ratios, the system reliably reaches consensus on the correct option ($\bar{\rho}_A \to 1$). In contrast, for large pooling errors and  $Q$, the population fails to coordinate on option $A$, resulting in $0 < \bar{\rho}_A  < 1$ and leading to a coexistence of opinion $A$ and $B$. Notably, the qualitative structure of both the consensus landscape and the decision-time patterns is almost the same for random and scale-free heterogeneous hypergraphs, indicating that the macroscopic behavior is driven mainly by the task parameters rather than by the specific features of the hypergraph topology.

A comparison between Fig.~\ref{fig::ABM2drdsf}, which depict the dynamics on homogeneous (d-uniform) hypergraphs, and Fig.~\ref{fig::ABMhetero_a}(a,b), illustrating the heterogeneous case, reveals that the overall qualitative behavior of the system is largely preserved when introducing heterogeneity in hyperedge sizes. In other words, the main dynamic regimes observed in the homogeneous setting—such as regions dominated by a particular opinion and transition zones—remain similar in the heterogeneous scenario. However, a noticeable difference emerges when closely comparing small $d-$uniform cases (notably panels~\ref{fig::ABM2drdsf}(a, e) and~\ref{fig::ABM2drdsf}(b, f)) with the heterogeneous case: the parameter region leading to convergence toward the higher-quality option is somewhat reduced under heterogeneity. This contraction can be interpreted as the effect of variability in group interaction sizes, which introduces fluctuations in local dynamics and tends to limit the efficiency of collective selection. Conversely, as the hyperedge size increases in the $d-$uniform case, the outcomes gradually approach those observed in the heterogeneous setting, suggesting that larger uniform interaction groups produce macroscopic dynamics similar to those induced by heterogeneity in hyperedge sizes.

\section{Conclusion}
Understanding how collective decisions emerge from individual interactions is one of the central challenge in the study of complex systems. To address this question, opinion dynamics provides a powerful theoretical framework, bridging social influence, information processing, and nonlinear dynamics. In this context, higher-order interactions, (i.e., those involving simultaneously groups rather than pairs) offer a richer and more realistic description of social processes, yet their role in shaping consensus formation and decision accuracy remains only partially understood.

In this work, we have proposed and analyzed an opinion dynamics model defined on $d$-uniform random and scale-free hypergraphs, in which social influence emerges from group-based (i.e., higher-order) rather than pairwise interactions. By introducing a Best-of-$2$ problem with two options, $A$ and $B$, with intrinsic qualities, $Q_A$ and $Q_B$ (with $Q_A>Q_B$) together with a pooling error $\alpha$, the model captures the interplay between information quality and collective uncertainty in a decision processes with higher-order interactions. Through a mean-field analysis, we proposed an ordinary differential equation that captures the dynamics of the Best-of-$2$ problem in a $d$-uniform hypergraph scenario. The analysis of this ODE helped us identify two critical values of the parameter $\alpha$, $\alpha_{\text{crit}}^{(1)}$ (characteristic of consensus to option $B$, i.e., the one associated to the worst option) and $\alpha_{\text{crit}}^{(2)}$ (characteristic of consensus on option $A$, i.e., the best one), which separate distinct regimes of consensus stability.
These analytical predictions were corroborated by extensive agent-based simulations on both $d-$uniform random and scale-free hypergraphs, and were found to remain valid even when considering heterogeneous hypergraphs in which the hyperedge sizes vary. Moreover, another key outcome of this analysis was that the bifurcation structure and critical thresholds ($\alpha_{\text{crit}}^{(1)}$ and $\alpha_{\text{crit}}^{(2)}$) are \textit{independent} of the underlying network topology (scale-free or random $d$-uniform hypergraph), depending solely on the interaction order $d$ and the opinion quality ratio $Q = Q_A / Q_B$. This finding points to a form of structural universality in higher-order interactions in opinion dynamics: collective transitions are governed primarily by intrinsic parameters rather than by the specific details of the network structure. Such robustness suggests that the fundamental mechanisms underlying consensus formation and decision accuracy may be largely insensitive to microscopic connectivity patterns when interactions occur within groups.

Considering beyond opinion dynamics, the proposed framework offers a general analytical foundation for studying collective decision-making in the Best-of-$n$ problems in systems governed by higher-order interactions. One of our future plan is to extend this approach to a Best-of-$n$ problem with $n>2$ scenarios and also in temporally evolving hypergraphs, and explore how stochasticity and adaptive feedback affect the bifurcation landscape. Another promising research direction is to establish a deeper link between the pooling error $\alpha$ and the group size $d$ in higher-order interactions. Previous investigations~\cite{reina2024speed} show that when $\alpha=0$, the dynamics correspond to a majority model; when $\alpha=1$, to a voter model; and when $\alpha \gg 1$, to a random choice process. This relationship could be formalized by setting $\alpha = d - 1$, which naturally connects the degree of information aggregation within a hyperedge to the behavioral regime of the agents. In particular, as $d>>1$, limited resources prevent agents from sampling all teammates, so that $\alpha \to +\infty$ effectively corresponds to a random decision process. Exploring this mapping in more realistic or heterogeneous systems could provide new insights into how the structure and size of interacting groups influence collective decision outcomes, particularly under conditions where excessive information leads to effective randomness—capturing the notion that “too much information kills the information.”\\

\section*{Acknowledgement}
We would like to warmly thank Andreagiovanni Reina for the insightful discussions that supported the early stages of this project. TN acknowledges support from the ``Reconstruction, Resilience and Recovery of Socio-Economic Networks'' RECON-NET - EP\_FAIR\_005 - PE0000013 ``FAIR'' - PNRR M4C2 Investment 1.3, financed by the European Union
– NextGenerationEU.

\appendix 
\section{The mean field model} \label{sec:app1}
For the sake of clarity and analytical tractability, Eq.~\ref{eq:dakdt} introduced in Sec.~\ref{sec:HHMF} can, after simple  manipulation, be decomposed into four distinct terms. This decomposition allows certain simplifications that lead to a more compact and interpretable form of the equation.
\begin{widetext} 
\begin{eqnarray}
\frac{da_k}{dt} &=& \sum _{r=1}^kg_k(d_r)\sum_{j_1,\dots,j_{d_r}}q_{j_1}\dots q_{j_{d_r}} \sum_{\omega=0}^{d_r} \pi_{d_r,\omega}P_\alpha\left( \frac{d_r-\omega}{d_r-\omega +\omega Q}\right) \notag \\
&-&a_k\sum _{r=1}^kg_k(d_r)\sum_{j_1,\dots,j_{d_r}}q_{j_1}\dots q_{j_{d_r}} \sum_{\omega=0}^{d_r} \pi_{d_r,\omega}\, .
\label{eq:hmf:dak}
\end{eqnarray}
\end{widetext} 
We trivially have $\sum_{\omega=0}^{d_r} \pi_{d_r,\omega}=1$ and, by assuming absence of correlations among nodes hyperdegrees, we also have $\sum_{j_1,\dots,j_{d_r}}q_{j_1}\dots q_{j_{d_r}}=\sum_{j_1} q_{j_1}\dots \sum_{j_{d_r}}q_{j_{d_r}}=1$ (A more detailed discussion is provided in Ref.~\cite{reina2024speed}). Under the simplifying assumption of a $d$-uniform hypergraph, i.e., where all hyperedges have the same size and thus $g_k(d_r)$ is uniform for all hyperedges, we have $\sum_{r=1}^{k} g_k(d_r) = 1$. This allows Eq.~\eqref{eq:hmf:dak} to be simplified as:
\begin{widetext} 
\begin{equation}
\frac{da_k}{dt} = \sum _{r=1}^kg_k(d_r)\sum_{j_1,\dots,j_{d_r}}q_{j_1}\dots q_{j_{d_r}} \sum_{\omega=0}^{d_r} \pi_{d_r,\omega}P_\alpha\left( \frac{d_r-\omega}{d_r-\omega +\omega Q}\right)-a_k\, .
\label{eq:dakdt2app}
\end{equation}
\end{widetext} 
Let us define $\langle a\rangle := \sum_j q_j a_{j+1}$, representing the average value of the fraction of agents holding opinion $A$ weighted by the probabilities $q_j$ of the corresponding states.  By using combinatorial arguments and assuming probability independence, we can show that 
\begin{equation*}
 \sum_{j_1,\dots,j_{d_r}}q_{j_1}\dots q_{j_{d_r}}  \pi_{{d_r},\omega} =  \binom{{d_r}}{\omega} \langle a\rangle^{{d_r}-\omega}\left(1-\langle a\rangle\right)^\omega\,;
\end{equation*}
the rough idea is that in $\pi_{{d_r},\omega}$ there are $({d_r}-\omega)$ events with probability $a_{j_m+1}$, thus $\omega$ with $(1-a_{j_m+1})$, and the binomial coefficient computes all possible permutations. We can thus rewrite Eq.\,\ref{eq:dakdt2app} as
\begin{widetext}
\begin{equation}
\label{eq:dakdt3}
\frac{da_k}{dt} =-a_k+ \sum _{r=1}^kg_k(d_r) \sum_{\omega=0}^{d_r}\binom{d_r}{\omega} \langle a\rangle^{d_r-\omega}\left(1-\langle a\rangle\right)^\omega P_\alpha\left( \frac{d_r-\omega}{d_r-\omega +\omega Q}\right)\, .
\end{equation}
\end{widetext}
By rewriting the previous equation with $k\rightarrow k+1$, by multiplying both sides by $q_{k}$ and by summing over $k$ to bring out $\langle a\rangle$ we get 
\begin{widetext}
\begin{eqnarray}
\label{eq:mainedo}
\frac{d\langle a \rangle}{dt}&=&\sum_k q_k \frac{da_{k+1}}{dt} = -\sum_k q_k a_{k+1}+\sum_k q_k \sum _{r=1}^{k+1}g_{k+1}(d_r) \sum_{\omega=0}^{d_r} \binom{d_r}{\omega} \langle a\rangle^{d_r-\omega}\left(1-\langle a\rangle\right)^\omega P_\alpha\left( \frac{d_r-\omega}{d_r-\omega +\omega Q}\right)\notag \\
&=&-\langle a \rangle+\sum_k q_k \sum _{r=1}^{k+1}g_{k+1}(d_r) \sum_{\omega=0}^{d_r} \binom{d_r}{\omega} \langle a\rangle^{d_r-\omega}\left(1-\langle a\rangle\right)^\omega P_\alpha\left( \frac{d_r-\omega}{d_r-\omega +\omega Q}\right)\equiv f_\alpha(\langle a\rangle)\, .
\end{eqnarray}
\end{widetext}

\section{Construction of the hypergraphs}\label{sec: appp2}
\subsection{$d$-uniform scale-free hypergraph}\label{sec: appps1}

Consider a random, undirected, $d$-uniform hypergraph with heterogeneous node participation, defined on a set of nodes $V=\{1,2,\dots,N\}$ and a set of hyperedges $E$. We assume that each hyperedge connects exactly $d$ distinct nodes. The principle of constructing this scale-free hypergraph is based on the model proposed by Jhun et al.~\cite{jhun2019simplicial}. To introduce structural heterogeneity, each node $i \in V$ is assigned a weight defined by Eq.\ref{eq:hetsf}:
\begin{equation}
\pi_i = \frac{i^{-\mu}}{\sum_{j=1}^{N} j^{-\mu}},
\label{eq:hetsf}
\end{equation}
where $0 < \mu < 1$ controls the degree of heterogeneity. This probability satisfy the normalization condition $\sum_{i=1}^{N}\pi_i=1$. This Eq.\ref{eq:hetsf} state that for $\mu=0$, all nodes are selected with equal probability, while increasing values of $\mu$ bias the selection toward high-weight nodes, producing broad hyperdegree distributions.

Hypergraph is constructed by generating $L = NK$ hyperedges, where $K$ is a density parameter. Each hyperedge is generated independently according to the following procedure:\\
i) A set of $d$ distinct nodes are sampled without replacement from $V$ according to the probability distribution $\{\pi_i\}$. The resulting set of nodes defines a hyperedge.\\
ii) Repeat i) $NK$ times. Multiple hyperedges are forbidden and duplicate realizations are discarded.\\
The hyperdegree follows a power-law distribution
\begin{equation}
    P_k \sim k^{-\gamma},
 \label{eq:eqpd}   
\end{equation}
with $\gamma = 1 + \frac{1}{\mu}$. \\
A scale-free regime is obtained for $2 < \gamma < 3$, corresponding to $0.5 < \mu < 1$, independently of $d$.
\subsubsection{Hyperdegree distribution}\label{sec:apphd}
In a $d$-uniform hypergraph, we define the hyperdegree $k_i$ of node $i$ as the number of hyperedges incident to that node. The hyperdegree distribution $p_k$ is obtained by counting $N_k$, the number of nodes with hyperdegree $k$, and normalizing by the total number of nodes $N$:
\[
p_k = \frac{N_k}{N} \sim k^{-\gamma}.
\]
For a scale-free hypergraph, this distribution follows a power-law form, as given in Eq.~\ref{eq:eqpd}.\\
To illustrate the method, we apply the procedure described in Sec.~\ref{sec: appps1} to build a network with $N = 10^5$ nodes, a hyperedge density $K = 1.3$ (i.e., a total of $NK$ hyperedges), a parameter $\mu = 0.65$ corresponding to $\gamma \approx 2.54$, and $d = 2$ (i.e., 3 nodes per hyperedge), which defines a $d$-uniform scale-free network. The resulting hyperdegree distribution $P_k$ is plotted in Fig.~\ref{fig::degreedis}(b) on a log-log scale and it verify the power-law behavior, typical of scale-free hypergraphs.

\subsection{$d$-uniform random hypergraph}\label{sec: appps2}

\begin{figure}[htp!]
\centering
\begin{tabular}{cc}
\includegraphics[width=0.25\textwidth]{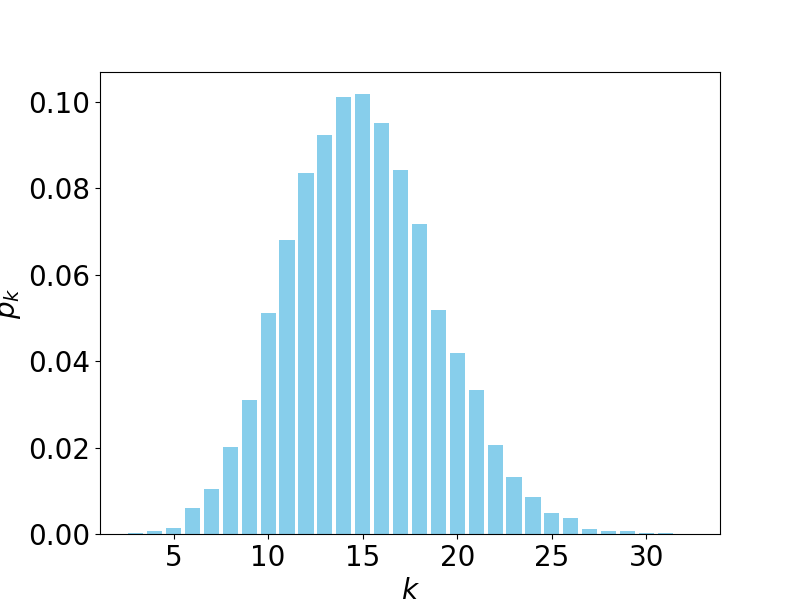}
&
\includegraphics[width=0.25\textwidth]{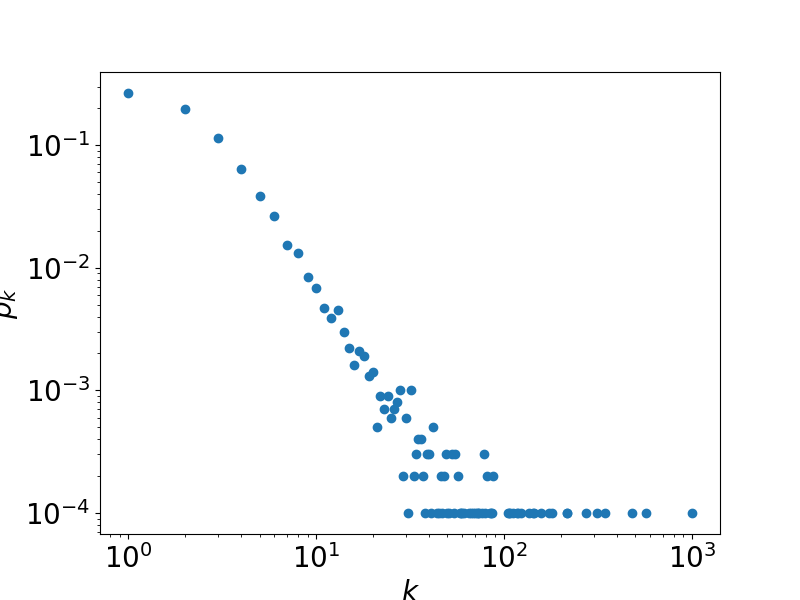}\\
(a) & (b)  \\
\end{tabular}
\caption{Hyperdegree distribution of a $d$-uniform hypergraph with $N = 10^5$ nodes and $d = 2$ (i.e., 3 nodes per hyperedge) at hyperedge density $K = 1.3$: (a) random hypergraph and (b) scale-free hypergraph with $\mu = 0.65$, corresponding to $\gamma \approx 2.54$.}
\label{fig::degreedis}
\end{figure}

To construct a $d$-uniform hypergraph with $N$ nodes and $NK$ hyperedges, where $K$ is the density of the hyperedges, we follow the procedure:\\
i) Select randomly a set of $d$ distinct nodes and the resulting set of nodes defines a hyperedge.\\
ii) Repeat i) $NK$ times, with multiple hyperedges forbidden and duplicate realizations discarded.\\
This procedure ensures that each hyperedge contains exactly $d$ distinct nodes and that all hyperedges are unique, producing a uniformly random $d$-uniform hypergraph.\\
An illustration of the hyperdegree distribution computed by using the procedure defined in Appendix~\ref{sec:apphd} is shown for a random $d$-uniform hypergraph with $d=2$, $N = 10^5$ nodes, and $K = 1.3$. The resulting hyperdegree distribution shown in Fig.~\ref{fig::degreedis}(a) follows a Poisson law, which is characteristic of random networks.

This method to generate $d$-uniform hypergraphs can be naturally generalized to heterogeneous, or non-uniform, hypergraphs. In this case, instead of fixing a unique hyperedge size $d$, step i) is modified by introducing an interval $[d_{\min},d_{\max}]$, from which the size of each hyperedge is sampled.

\nocite{*}

\end{document}